
\input phyzzx
\let\text\hbox

\normalspace
\def\delt{\partial_t}
\def\delx{\partial_x}
\def\dely{\partial_y}
\def\delz{\partial_z}
\def\delq{\partial_q}
\def\delqp{\partial_{q'}}

\def\adelx{{|\partial_x|}}

\def\half{{1 \over 2}}
\def\pv{{\int\hskip-12pt-\,}}

\FRONTPAGE
\title{{\bf
    A CONTINUUM DESCRIPTION OF \\
    SUPER-CALOGERO MODELS  \\
     }}
\bigskip
\centerline{Andr\'e Johann van Tonder}
\bigskip
\centerline{Department of Physics, University of the Witwatersrand}
\centerline{PO WITS 2050, South Africa}

\vfill
\noindent
A thesis submitted to the Faculty of Science, University of the
Witwatersrand, Johannesburg, in fulfilment of the requirements of the
degree of Doctor of Philosophy.

\vfill
\noindent
\line {CNLS-91-01 \hfill}
\line {March 1992 \hfill}
\line {Bitnet address: 005CNLS@witsvma.wits.ac.za \hfill}
\endpage

\centerline {\bf ABSTRACT}

\bigskip

\sl
\noindent
In this thesis an investigation is made of the super-Calogero model
with particular emphasis on its continuum formulation and possible
application in the context of supersymmetrizing the bosonic collective
$d=1$ string field theory.

We start with a discussion of the model both in a discrete and
continuum formulation, demonstrating its equivalence to the
Marinari-Parisi supersymmetric matrix model and the
Jevicki-Rodrigues supersymmetrized collective field theory. Upon
quantization,
the continuum fields are found to have nontrivial commutation
relations
involving square roots of the density field, leading to an infinite
sequence of higher order vertices in the perturbative Hamiltonian as
well as the supersymmetry generators.

We then discuss the potential free case, with an explicit
calculation of the spectrum and the cubic vertices.
After comparing a particular spacetime
formulation of the theory with gauge fixed two-dimensional
supergravity,
the exact spectrum, as previously obtained in the
bosonic collective field theory, is generalised to the
supersymmetric
case.
In addition to this, a formulation is
postulated in which to investigate nonperturbative effects such as
solitons in a semiclassical analysis.

For the harmonic case, the semiclassical spectrum
and cubic vertices are
then calculated, an undertaking that turns out to be very sensitive
to the method of regularization.
If one wants to preserve supersymmetry, it seems that one cannot
generate a bosonic sector compatible with the bosonic collective
string field theory.

We then move on to more general potentials.  After discussing
the semiclassical spectrum,  we conclude with an investigation of
the scaling properties of the
superpotentials and the possibility of a spacetime interpretation of
the theory.  It is argued that
the scaling poperties of the vacuum configuration
correspond to that found in zero-dimensional $c<1$ models, in contrast
to the $c=1$ behaviour of the bosonic collective field theory.

\rm

\vfill
\endpage

\centerline {\bf DECLARATION}

\bigskip

\noindent
I declare that this thesis is my own, unaided work.  It is being
submitted for the degree of Doctor of Philosophy at the University of
the Witwatersrand, Johannesburg.  It has not been submitted before for
any degree or examination at any other University.

\bigskip
\bigskip

\noindent
$\overline {\text {Andr\'e van Tonder}}$

\bigskip
\noindent
$\underline{\qquad\qquad}$ day of
$\underline{\qquad\qquad\qquad}$, 1992

\vfill
\endpage

\centerline {\bf ACKNOWLEDGEMENTS}

\bigskip

\noindent
I am deeply indebted to Jo\~ao Rodrigues for his excellent supervision
of my work and many discussions.  I would also like to express my
gratitude to the theory group at Brown University for their
hospitality during a visit of mine in the final months of my
doctorate, and to thank Antal Jevicki, Branislav Sazdovi\'c,
Kre\v simir Demeterfi, Katsumi Itoh and Branko Uro\u sevi\'c for
stimulating discussions on matters related to this work.

Ek wil ook my opregte waardering uitspreek teenoor my gesin, op wie se
liefdevolle ondersteuning ek altyd kan staat maak.

\vfill
\endpage

\centerline {\bf CONTENTS}

\bigskip

\noindent
\line{{\bf Chapter one.} \hfill 7}
\line{Introduction\hfill}
\line{{\bf Chapter two.} \hfill 9}
\line{The Supersymmetric Model \hfill}
\line{$\qquad\qquad$ 2.1. Motivation \hfill 9}
\line{$\qquad\qquad$ 2.2. The super-Calogero model \hfill 10}
\line{$\qquad\qquad$ 2.3. Continuum description \hfill 12}
\line{$\qquad\qquad$ 2.4. Supersymmetrized collective field model \hfill
14}
\line{$\qquad\qquad$ 2.5. Supersymmetric matrix model \hfill 18}
\line{$\qquad\qquad$ 2.6. Explicit $N$-dependence \hfill 22}
\line{{\bf Chapter three.} \hfill 25}
\line{Perturbation Theory \hfill}
\line{{\bf Chapter four.} \hfill 31}
\line{The Potential Free Case \hfill}
\line{$\qquad\qquad$ 4.1. Motivation \hfill 31}
\line{$\qquad\qquad$ 4.2. The vacuum density \hfill 31}
\line{$\qquad\qquad$ 4.3. Perturbative expansion \hfill 35}
\line{$\qquad\qquad$ 4.4. Comparison with supergravity \hfill 37}
\line{$\qquad\qquad$ 4.5. Exact nonperturbative results \hfill 43}
\line{$\qquad\qquad$ 4.6. Alternative representation of exact states \hfill
44}
\line{$\qquad\qquad$ 4.7. Semiclassical analysis \hfill 47}
\line{{\bf Chapter five.} \hfill 50}
\line{The Harmonic Case \hfill}
\line{$\qquad\qquad$ 5.1. Motivation \hfill 50}
\line{$\qquad\qquad$ 5.2. The vacuum density \hfill 50}
\line{$\qquad\qquad$ 5.3. The quadratic spectrum \hfill 52}
\line{$\qquad\qquad$ 5.4. Three point functions \hfill 55}
\line{$\qquad\qquad$ 5.5. Turning point divergences \hfill 58}
\line{{\bf Chapter six.} \hfill 64}
\line{General Potential \hfill}
\line{$\qquad\qquad$ 6.1. Semiclassical spectrum \hfill 64}
\line{$\qquad\qquad$ 6.2. Scaling limit \hfill 70}
\line{{\bf Chapter seven.} \hfill 75}
\line{Conclusions \hfill}
\line{{\bf References.} \hfill 77}

\vfill
\endpage

\chapter{INTRODUCTION}

\noindent
{}From a continuum description of the eigenvalue dynamics of matrix models
[1], the collective field method has emerged as a field theory of
one dimensional bosonic strings [2]
and has provided much insight into their physics.

At the same time, the fact that the eigenvalue dynamics is
described by an exactly
solvable $N$-particle model of a type that was extensively studied
by Calogero and Perelomov [34] in the sixties and seventies has
proven very useful in the formulation and resolution of issues of
symmetry and integrability, both in the discrete and in the collective
field language [4].

All indications are that the collective string field theory
is a complete theory, in that perturbative calculations [6, 7]
are in agreement with matrix model [8], first quantized [9]
and exact fermionic [10, 11] calculations.

It is therefore of great interest to develop a field theory of
$d=1$ superstrings.

The study of non critical superstrings [12] has centered, essentially,
on two approaches:
one is based on super Liouville,
first quantized calculations, first for $d < 1$ [14] and
recently for the $d=1$ superstring [15]
and the other is based on the
supersymmetrization of the eigenvalue dynamics
of matrix models.
The latter has been done both in the discrete language [13]
and directly in the continuum, by supersymmetrizing the bosonic
collective field theory [16].

In both approaches one encounters problems of
interpretation.  In the Liouville language it is unclear whether
spacetime supersymmetry exists.  In the matrix model the
spacetime interpretation itself is problematic and it is not known at
present whether one is indeed describing a theory of one dimensional
superstrings.

While the bosonic matrix model corresponds to the exactly solvable
Calogero model, its supersymmetrization corresponds to
the super-Calogero model as introduced in reference [23], or a
suitable continuum approximation of it.

It is the purpose of this thesis to study the super-Calogero model
with particular emphasis on its continuum formulation
and possible application in the context
of supersymmetrizing the bosonic collective $d=1$ string field theory.

The text is organized as follows.  In chapter 2, after a brief
historical introduction, we give the continuum formulation of the
super-Calogero model and show its equivalence to both
the Marinari-Parisi supermatrix model and the Jevicki-Rodrigues
supersymmetric  collective field theory.
In chapter 3 we set up the perturbation theory for a general
superpotential.  In chapter 4 we specialize to the potential free case
and extend some exact nonperturbative results from the bosonic
collective field
theory to the supersymmetric case.  In chapter 5 we discuss the case
of a harmonic superpotential, with special reference to issues of
regularization. In chapter 6 we return to the general potential
with a discussion of the spectrum and the scaling limit.
The last chapter is reserved for conclusions and outlook.

\endpage

\chapter{THE SUPERSYMMETRIC MODEL}

\section {Motivation}

\noindent
The bosonic Calogero model was first studied as an example of an
exactly
solvable $N$-particle quantum mechanical system by Calogero and
Perelomov [34] in the sixties and seventies.  Its Hamiltonian
in the centre of mass frame of reference is given by
$$
  H_B = \half\sum_{i=1}^N \, p_i^2 + V_B(x_1,\ldots,x_N),
  \eqn \bCalogeroA
$$
where the potential $V_B$ is chosen to be
$$
  V_B(x_1,\ldots,x_N) = {{\omega^2}\over 2}\sum_i x_i^2
     + {\varepsilon^2\over 2}\sum_{i\ne j}{1\over{(x_i - x_j)^2}}.
\eqn \bCalogeroB $$

We see that the potential consists of a harmonic piece as well as a
term describing
a repulsive interparticle force.

It has been known for some time [8, 18] that at the quantum level the
dynamics of the singlet
sector of certain one dimensional matrix models are described by
systems of Calogero
type.  Indeed, when one quantizes the singlet sector, parametrized by
the eigenvalues, of an $SU(N)$ invariant matrix model, there appears a
Jacobian
that can be reinterpreted as an effective interaction between the
eigenvalues.  This effective interaction
is identical to the singular term in \bCalogeroB.

Even at the classical level one can obtain this effective repulsive
interaction by imposing specific constraints on the conserved angular
momenta corresponding to the $SU(N)$ symmetry.  This fact was noted in
[21] in the context of describing the planar limit of quantized
matrix models.

Recently systems of Calogero type have attracted renewed
interest due to the fact that they appear in the matrix model approach
to one-dimensional bosonic quantum string field theory.
A stringy continuum limit is obtained via a double scaling in
which one
essentially replaces the harmonic part of
the potential \bCalogeroB\ with an inverted harmonic interaction,
letting $N\rightarrow\infty$ while the fermi level is taken to
approach the maximum of the inverted potential.

There have been attempts, initiated by Marinari and Parisi [13], to
define a supersymmetric string field theory in one
dimension via a matrix model formulated in a
$(1,1)$-dimensional
superspace.
On the other hand, existing results for the Calogero system have been
extended by Freedman and Mende to its natural supersymmetrization, to
be discussed in the next section.
This super-Calogero
system
is essentially equivalent to the Marinari-Parisi matrix model, as we
shall demonstrate in section (2.5).

A collective field description of the matrix
model formulation of $d=1$ bosonic string theory was given for the
first time by Das and Jevicki in [2].  In this approach, the string
field theory was seen to correspond
to a cubic field theory of a massless tachyon particle in two
dimensions.
This bosonic collective field theory was supersymmetrized by Jevicki
and Rodrigues in [16] for the noninverted harmonic oscillator. As we
shall see in section (2.4), the resulting theory
is equivalent to the super-Calogero model in a continuum formulation
to be established in section (2.3).

\section {The super-Calogero model}

\noindent
The supersymmetric generalization of the Calogero model was first
investigated by Freedman and Mende in [23].
For completeness, we repeat their construction here.
Using the approach of Witten [29] one introduces, in addition to the
bosonic coordinates $x_i$, the fermionic coordinates
$\psi_i$ and $\psi_i^\dagger$ satisfying the standard anti-commutation
relations
$\{\psi_i,\psi_j\}=0$, $\{\psi_i^\dagger,\psi_j^\dagger\}=0$ and
$\{\psi_i,\psi_j^\dagger\}=\delta_{ij}$.  One then defines
supercharges in terms of a so-called superpotential
$W(x_i,\ldots,x_N)$ as
$$
 \eqalign{
  Q &\equiv \sum_i \psi_i^\dagger\left(
            p_i - i\, {{\partial W}\over{\partial x_i}}\right),  \cr
  Q^\dagger &\equiv \sum_i \psi_i\left(
            p_i + i\, {{\partial W}\over{\partial x_i}}\right)  \cr
 }  \eqn \sCalogeroA
$$
satisfying $\{Q,Q\} = 0 = \{Q^\dagger,Q^\dagger\}$.  The Hamiltonian
is constructed as
$$
 \eqalign{
  H_S &\equiv \half \{  Q, Q^\dagger   \}   \cr
      &=  \half\sum_i \left(p_i^2
             + \left(
            {{\partial W}\over{\partial x_i}} \right)^2 \right)
      +\half\sum_{i,j}\, [\psi^\dagger_i,\psi_j]\,
                {{\partial^2 W}\over{\partial x_i \partial x_j}}
 }  \eqn \sCalogeroB
$$
and commutes with $Q$ and $Q^\dagger$.

Choosing, for example,
$$
  W(x_1,\ldots,x_N) = {\omega\over{2}}\sum_{i}x_i^2
              + {\varepsilon\over 2}\sum_{i\ne j}\log |x_i-x_j|,
\eqn \CalogeroW
$$
one finds, after some algebra [23],
$$
 \eqalign{
  H_S &= \half\sum_{i} \, p_i^2 +
          {{\omega^2}\over{2}}\sum_{i}x_i^2
           + {{\varepsilon^2}\over 2}\sum_{i\ne j}{1\over{(x_i -
            x_j)^2}} \cr
      &+ \omega \sum_i \psi_i^\dagger\psi_i
       - {\varepsilon\over 2}\sum_{i \ne j}\,
       [\psi_i^\dagger-\psi_j^\dagger,\psi_i]\,{1\over{(x_i-x_j)^2}}
         \cr
     &- {\omega\over 2}\, (1 - \varepsilon N) (N-1). \cr
 }  \eqn \sCalogeroH
$$
The bosonic part of this supersymmetric Hamiltonian
coincides, apart from an additive constant, with the original Calogero
model \bCalogeroA\ and \bCalogeroB.

The model \sCalogeroH\ has the same form as the one studied by
Freedman
and Mende in the centre of mass frame of reference.  However, their
coordinates are centre of mass coordinates and therefore satisfy the
constraints
$$
  \sum_i x_i = \sum_i p_i = \sum_i \psi^\dagger_i = \sum_i \psi_i =
0, $$
which are second class and modify the canonical brackets.
However, as we are interested in
the correspondence of the above system with matrix models
we shall
ignore the constraints and assume standard commutators.

This concludes the discussion of the supersymmetric generalization of
the Calogero model.  We now move on to a formulation of the model in
terms of continuum fields.

\section{Continuum description}

\noindent
We now follow the lead of [21] and
[16]
to set up a continuum approximation of the discrete model \sCalogeroA\
and \sCalogeroB.
This approximation is assumed to become exact in the limit
$N\to\infty$, where the discrete distribution $x_i$ will
approximate a continuous density.  With this in mind, we introduce
the continuum index $x$ and define the fields
$$
  \phi(x) \equiv \delx\varphi \equiv \sum_i \delta(x-x_i),
\qquad
  \phi\sigma(x)\equiv -\sum_i\delta(x-x_i)\,p_i,
$$
$$
  \psi(x) = -\sum_i \delta(x-x_i)\,\psi_i, \qquad
  \psi^\dagger(x) = -\sum_i \delta(x-x_i)\,\psi^\dagger_i.
  \eqn \DiscToCont
$$
These fields satisfy commutation relations
$$
 \eqalign{
  [\sigma(x), \varphi(y)] &= -i\, \delta (x-y),  \cr
  [\sigma(x), \psi^\dagger(y)] &= i\, {{\psi^\dagger}\over\phi}(x)\,
                          \delx\delta(x-y),   \cr
  [\sigma(x), \psi(y)] &= i\, {{\psi}\over\phi}(x)\,
                          \delx\delta(x-y),   \cr
  \{\psi(x), \psi^\dagger(y)\} &= \phi(x)\,\delta(x-y).   \cr
 } \eqn \cCommutations
$$
These equalities are easily proved
using the identity
$$\delta(x-x_i)f(x) =
\delta(x-x_i)f(x_i),   \eqno\eq$$
 where $f$ is an arbitrary function.

Let us now rewrite the Calogero supercharges \sCalogeroA\ in terms
of the continuum fields.  At the classical level, using $\int dx\,
{{\delta(x-x_i)\,\delta(x-x_j)}/{\phi(x)}} = \delta_{ij}$, which
can be
established via \?, and the fact that by the chain rule
$\partial
W/\partial{x_i} = \delta W/\delta\varphi(x_i)$,
it is easy to show that the
supercharges can equivalently be expressed as
$$
 \eqalign{
  Q &= \int dx\,\psi^\dagger(x)\,(\sigma(x) -i\,W_{;x}),  \cr
  Q^\dagger &= \int dx\,\psi(x)\,(\sigma(x) +i\,W_{;x}),  \cr
 } \eqn \continuumQ
$$
where we have used the notation $W_{;x}\equiv\delta W/\delta\varphi(x)$.

In a careful quantum mechanical treatment of bosonic matrix
models, additional terms arise in the corresponding
continuum Hamiltonian [1, 37].
However, leaving out these terms, perturbative calculations of
scattering amplitudes [15]
reproduce exactly those obtained in the continuum approach.  In the
free potential case [17] an exact solution exhibits two
single particle branches, one of which can always be described
semiclassically by the inclusion of the extra terms.
We assume that a similar
mechanism would apply here and
postulate \continuumQ\ to be the full quantum mechanical
supercharges.

To find the equivalent expression for the Hamiltonian \sCalogeroB\ in
terms
of the continuum fields, we can take advantage of the fact that the
corresponding supercharges are equivalent and use \continuumQ\ to
construct the continuum Hamiltonian. We find, using
$H=\half\{Q,Q^\dagger\}$, that
$$
\eqalign{
  H &= \half\int dx\, \phi\,\sigma^2 +\half\int dx\,
       \phi\,(W_{;x})^2  \cr
    &- \half\int {{dx}\over\phi}\,[\psi^\dagger,\psi]\,\delx W_{;x}
      + \half\int dx \int dy\, [\psi^\dagger(x),\psi(y)]\,W_{;xy}.
}  \eqn \continuumH
$$
{}From the definition of the field $\phi$, we see that the continuum
Hamiltonian must be accompanied by the constraint
$$
  \int dx\, \phi(x) = N.
$$

To summarise, in this section we have set up a continuum formulation
of the super-Calogero model, based on approximating discrete
eigenvalue densities by continuous functions.  This approximation is
expected to become exact once a suitable $N\to\infty$ limit is taken.
We now have a language in which to compare the Calogero model with
the Jevicki-Rodrigues supersymmetrization of the bosonic collective
field theory, which we do in the next section.

\section {Supersymmetrized collective field
model}

\noindent
In [16] a supersymmetric extension of the bosonic collective field
theory was described.
This was done by noting
that the collective field theory can be seen as a metric
theory, which can be supersymmetrized via
a standard procedure.

To see how this works, observe that the kinetic term of the bosonic
collective Lagrangian
$$
  L = \half\int {{dx}\over\phi}\, \dot\varphi^2 -
            {{\pi^2}\over 6}\int dx\,\phi^3 - \int dx\,v\phi
  \eqno \eq
$$
can be written in the form
$$
 L_T = \half\int dx \int
dy\,\dot\varphi(x,t)\,g_{xy}\,\dot\varphi(y,t),
$$
where the continuous index metric is given by
$$
 g_{xy}(\varphi) = {1\over{\phi(x)}} \,\delta (x-y).
 \eqn \gxy
$$

A standard supersymmetrization of a theory of this type is given by
$$
 \eqalign{
  L &= \half (\dot q^a g_{ab}\, \dot q^b - g^{ab}\,\partial_a W\,
        \partial_b W) + i\,\psi^{\dagger a}g_{ab}\,\dot\psi^b   \cr
    &+ i\,\psi^{\dagger a}\Gamma_{bc,a}\,\dot q^c\psi^b
      + \half\,[\psi^{\dagger c},\psi^d]\,\Gamma^a_{cd}\,\partial_a W
      - \half\,[\psi^{\dagger a},\psi^b]\,\partial_a\partial_b W,
 }   \eqno \eq
$$
where $q^a$ are the bosonic variables, $g_{ab}$ is a metric, $W$ is a
superpotential and $\Gamma^c_{ab}$ are the Christoffel symbols.
This can be obtained by a classical point transformation $x^i \equiv
x^i(q^a)$, $\psi^i = (\partial x^i/\partial q^a)\psi^a \equiv
e^i_a\psi^a$ from the Lagrangian
$$
  L = \sum_{i=1}^N \left(\half\dot x_i^2 -\half(\partial_i W)^2\right)
    + i\sum_{i=1}^N \,\psi_i^\dagger\dot\psi_i
    -\sum_{ij}\,[\psi^\dagger_i,\psi_j]\,\partial_i\partial_j W,
$$
which is  a multidimensional generalization of one-dimensional
supersymmetric quantum mechanics [29].

Note the expressions of the form $\half\,[\psi^{\dagger},\psi]$ here.
Classically, taking $\phi$ and $\phi^\dagger$ to be Grassman
variables, we have the identity
$\half\,[\psi^{\dagger},\psi] =\psi^\dagger \psi$.  At the
quantum level, however,
this replacement would destroy supersymmetry unless an additional
term were added
to the bosonic potential in \?.  Thus \? is the proper Lagrangian to
use for quantization.

In our case the $q^a$ are replaced by $\varphi(x)$ and $g_{xy}$ is
given by \gxy. One obtains the Lagrangian
$$
 \eqalign{
  L &= \half \int{{dx} \over \phi}\, {\dot \varphi}^2
      - \half\int dx\,\phi\,(W_{;x})^2 \cr
     &+ {i} \int {{dx} \over \phi}\,\psi^\dagger\dot\psi
      + {i} \int{{dx} \over \phi}\,\dot\varphi\,
         \delx\left({{\psi^\dagger} \over \phi}\right) \psi
           \cr
     &+ \half\int{{dx} \over \phi}\, [\psi^\dagger,\psi] \,\delx
        W_{;x}
      - \half\int dx\int dy\, [\psi^\dagger(x),\psi(y)]\, W_{;xy}. \cr
 }
  \eqn \nonsymmetricL
$$
This is equivalent, via a partial integration, to
$$
 \eqalign{
  L &= \half \int{{dx} \over \phi}\, {\dot \varphi}^2
      - \half\int dx\,\phi\,(W_{;x})^2 \cr
     &+ {i \over 2} \int {{dx} \over \phi}\,(\psi^\dagger\dot\psi
             - {\dot \psi}^\dagger \psi)
      + {i \over 2} \int{{dx} \over \phi}\,\dot\varphi
         \left[\delx\left({{\psi^\dagger} \over \phi}\right) \psi
          -\psi^\dagger \delx \left({\psi \over\phi}\right)\right] \cr
     &+ \half\int{{dx} \over \phi}\, [\psi^\dagger, \psi] \,\delx
         W_{;x}
      - \half\int dx\int dy\, [\psi^\dagger(x),\psi(y)]\, W_{;xy}. \cr
 }
  \eqn \symmetricL
$$

Our motivation for this rewriting is that in the form \symmetricL\ the
momentum conjugate to the field $\varphi$ is manifestly hermitian,
while in \nonsymmetricL\ it is nonhermitian.  Even so, one would like
the respective theories to be equivalent at the quantum level.
Though it would take us too far afield to show it here, this indeed
turns out to be the case:  in both cases we can define a hermitian
$\sigma$ which satisfies the commutation relations \cCommutations\ and
in terms of which the Hamiltonian can be written in the form
\continuumH.

The conjugate momenta are given by
$$
     p(x)  = {{\partial L} \over {\partial\dot\varphi}}
           = {{\dot\varphi}\over\phi}
             + {i\over 2} \left[\delx\left({{\psi^\dagger} \over
              \phi}\right) {\psi\over\phi}
              -{{\psi^\dagger}\over\phi} \delx \left({\psi \over
                \phi}\right)\right],    \eqno \eq
$$
$$
     \Pi = {{\partial L} \over {\partial\dot\psi}}
         = {i\over 2} {{\psi^\dagger}\over\phi}, \qquad
     \Pi^\dagger = {{\partial L} \over {\partial\dot\psi^\dagger}}
         = -{i\over 2} {{\psi}\over\phi}. \eqno \eq
$$

Identifying the second class constraints
$$
 \eqalign{
    \chi &= \Pi - {i\over 2} {{\psi^\dagger}\over\phi}, \cr
    \bar\chi &= \Pi^\dagger + {i\over 2} {{\psi}\over\phi},  \cr
 } \eqno \eq
$$
one uses Dirac brackets [16, 26] to obtain the following equal
time brackets
$$
  [\varphi(x),\varphi(y)] = 0, \quad [\varphi(x),p(y)] = i\delta(x-y),
  \quad \{\psi(x),\psi^\dagger(y)\} = \phi(x)\delta(x-y), \eqn
\collCommA
$$
$$
  [\varphi(x),\psi(y)] = [\varphi(x),\psi^\dagger(y)] = 0,
  \eqn \collCommB
$$
$$
  [\,p(x),\psi(y)] = {i\over
   2}{{\psi(y)}\over{\phi(y)}}\,\delx\delta(x-y),
  \quad
  [\,p(x),\psi^\dagger(y)] = {i\over
     2}{{\psi^\dagger(y)}\over{\phi(y)}}\,\delx\delta(x-y). \eqn
\collCommC
$$
In terms of
$$
  \sigma(x) \equiv p(x)
              - {i\over 2} \left[\delx\left({{\psi^\dagger} \over
               \phi}\right) {\psi\over\phi}
               -{\psi^\dagger\over\phi} \delx \left({\psi \over
                 \phi}\right)\right]    \eqn \sigmaDef
$$
the Hamiltonian then takes the form
$$
\eqalign{
  H &= \half\int dx\, \sigma\phi\sigma +\half\int dx\,
       \phi\,(W_{;x})^2  \cr
    &- \half\int {{dx}\over\phi}\,[\psi^\dagger,\psi]\,\delx W_{;x}
      + \half\int dx \int dy\, [\psi^\dagger(x),\psi(y)]\,W_{;xy},
}  \eqno \eq
$$
while from equations \collCommA, \collCommC\ and \sigmaDef\ it
follows that
$$
 [\sigma(x),\psi(y)] = i\,{{\psi}\over{\phi}}(x)\,\delx\delta(x-y),
 \qquad
 [\sigma(x),\psi^\dagger(y)] = i\,
    {{\psi^\dagger}\over{\phi}}(x)\,\delx\delta(x-y).
\eqn\SigmaPsiComm $$

The collective field $\phi$ satisfies the constraint
$$
  \int dx\,\phi(x) = N.
$$

This Hamiltonian and these commutation relations are
identical with those derived in the continuum description of the
super-Calogero model in \cCommutations\ and \continuumH.

What is unusual about the commutations \SigmaPsiComm\ is of course the
fact
that the bosonic momentum does not commute with the fermionic fields.
We observe that from \collCommA\ and \collCommC\ we have
$$
 \biggl[ p(x), {{\psi(y)} \over {\sqrt{\phi(y)}}}\biggr] = 0,
 \qquad
 \biggl[ p(x), {{\psi^\dagger(y)} \over {\sqrt{\phi(y)}}}\biggr] = 0.
 \eqno \eq
$$
We therefore rescale $\psi(x)\to\sqrt{\phi(x)}\,\psi(x)$;
$\psi^\dagger(x)\to\sqrt{\phi(x)}\,\psi^\dagger(x)$ to obtain for the
Lagrangian
$$
 \eqalign{
  L &= \half \int{{dx} \over \phi}\, {\dot \varphi}^2
      - \half\int dx\,\phi\,(W_{;x})^2 \cr
     &+ {i \over 2} \int {{dx}}\,(\psi^\dagger\dot\psi
             - {\dot \psi}^\dagger \psi)
      + {i \over 2} \int{{dx} \over \phi}\,\dot\varphi
         \left[\delx{{\psi^\dagger}} \psi
          -\psi^\dagger \delx {\psi}\right] \cr
     &+ \half\int{{dx}}\, [\psi^\dagger,\psi] \,\delx W_{;x}
      - \half\int dx\int dy\,
        [\psi^\dagger(x),\psi(y)]\sqrt{\phi(x)}\, W_{;xy}\,
          \sqrt{\phi(y)}, \cr
 }
  \eqn \rescaledL
$$
for the Hamiltonian
$$
\eqalign{
  H &= \half\int {{dx}\over\phi}\,
        \left(\phi p - {i\over 2} \left[(\delx\psi^\dagger)\psi
           -\psi^\dagger (\delx \psi)\right]\right)^2 \cr
    &+\half\int dx\,
       \phi\,(W_{;x})^2
    - \half\int {{dx}}\,[\psi^\dagger, \psi]\,\delx W_{;x}   \cr
    &+ \half\int dx \int dy\,
[\psi^\dagger(x),\psi(y)]\sqrt{\phi(x)}\,W_{;xy}\,\sqrt{\phi(y)},\cr
} \eqn \rescaledH
$$
and for the supercharges
$$
 \eqalign{
  Q &= \int dx\,\psi^\dagger(x)\sqrt{\phi(x)}\,\left(
       {{p(x)}} - {i\over {2\phi}}
       \left((\delx\psi^\dagger)\psi-\psi^\dagger (\delx \psi)\right)
       -i\,W_{;x}\right),
\cr
  Q^\dagger &= \int dx\,\psi(x)\sqrt{\phi(x)}\,\left(
       {{p(x)}} - {i\over {2\phi}}
       \left((\delx\psi^\dagger)\psi-\psi^\dagger (\delx \psi)\right)
       +i\,W_{;x}\right),
\cr
 } \eqn \rescaledQ
$$
with standard commutators
$$
  [\varphi(x),\varphi(y)] = 0, \quad [\varphi(x),p(y)] =
       i\,\delta(x-y),
  \quad \{\psi(x),\psi^\dagger(y)\} = \delta(x-y),
$$
$$
  [\varphi(x),\psi(y)] = [\varphi(x),\psi^\dagger(y)] =
  [\,p(x),\psi(y)] =
  [\,p(x),\psi^\dagger(y)] = 0.  \eqno \eq
$$
The square root factors appearing in equations \rescaledL\ to
\rescaledQ, when
expanded around the large $N$ background configuration, will generate
an infinite perturbative expansion.

To summarise, in this section we showed this approach to
supersymmetrizing the bosonic collective field theory
to be a special case of
the super-Calogero model in the continuum formulation.  In the next
section we shall establish the correspondence of the
super-Calogero model with the supersymmetric matrix model.

\section {Supersymmetric matrix model}

\noindent
In this section we will demonstrate the equivalence of the
Marinari-Parisi supersymmetric matrix model [13] to the super-Calogero
model of section (2.2).
Our analysis
is based on that of Dabholkar in [13].

Consider a theory of $N\times N$ hermitean matrices on a
$(1,1)$-dimensional superspace with action
$$
  S = \int dt\,d\theta\,d\bar\theta\, \left(\half{\rm Tr}\, \left(
      \bar D \Phi\, D\Phi\right) + \bar W (\Phi)\right).
   \eqn \matrixS
$$
Here $D \equiv {\partial_{\bar\theta} - i\,\theta\,\partial_t}$, and
we can expand $\Phi$ as
$$
  \Phi \equiv  M + \Psi^\dagger \theta +
     \bar\theta\Psi + \bar\theta\theta F,
  \eqno \eq
$$
where $M$ and $F$ are hermitean, and where we have used the
conventions $(\alpha\beta)^* = \beta^*\alpha^*$ and
$(\partial_{\bar\theta})^* = -\partial_\theta$.

The Feynman diagrams of a matrix theory can be topologically
classified according to the genus.  If one takes ${\bar W}$ to be of
the form
$$
  \bar W (\Phi) = {\rm Tr}\left(
    g_2\,\Phi^2 + {{g_3}\over{\sqrt N}}\,\Phi^3 + \cdots
     + {{g_p}\over{N^{p/2 - 1}}}\,\Phi^p\right),
  \eqn \generalW
$$
then by a simple topological argument [8]
it follows
that a diagram of genus $\Gamma$ carries a factor $N^{2 - 2\Gamma}$
and can therefore naively be interpreted as a
supertriangulation
of the corresponding diagram in the perturbation expansion of a string
theory with coupling constant $1/N^2$.

Upon quantization one finds the Hamiltonian
$$
\eqalign{
  H &= \half\,{\rm Tr} \left(P^2 + {{\partial \bar W(
M)}\over{\partial
         M^*}}\,
       {{\partial \bar W( M)}\over{\partial M}}\right)
       + \sum_{ijkl}\, [\Psi^*_{ji}, \Psi_{kl}]\,
         {{\partial \bar W( M)}\over{\partial M^*_{ij}\,\partial
M_{kl}}}, }  \eqno \eq
$$
and supercharges
$$
\eqalign{
  Q         &= \sum_{ij} \Psi^*_{ij}\left( P^*_{ij} - i\,
                {{\partial \bar W( M)}\over{\partial M^*_{ij}}}
\right), \cr
  Q^\dagger &= \sum_{ij} \Psi_{ij}\left( P_{ij} + i\,
                {{\partial \bar W( M)}\over{\partial M_{ij}}} \right),
\cr
}  \eqn \matrixQ
$$
where $[P_{ij}, M_{kl}] = -i\,\delta_{ik}\delta_{jl}$ and
$\{\Psi_{ij},\Psi_{kl}\} = \delta_{ik}\delta_{jl}$.

In the bosonic matrix model one observes at this point that the model
is invariant under unitary transformations, which implies that one can
restrict the theory to the invariant subspace consisting of those
wavefunctions that only depend
on the eigenvalues.

An analogous procedure can be followed in the
supersymmetric case:
Let $U$ be the unitary transformation that diagonalizes the bosonic
component $ M$ of the matrix $\Phi$.  In general the
fermionic components $\Psi$ and $\Psi^\dagger$ of
$\Phi$ will not be diagonalized by $U$.  Nevertheless, writing
the diagonal elements as
$$
  (U\Phi U^\dagger)_{ii}
   \equiv \lambda_i + \bar\theta\,\psi_i + \psi_i^\dagger\,\theta +
   \bar\theta\theta f_i,     \eqno \eq
$$
we shall see that we can define an invariant subspace
consisting of those wave functions that depend
only on $(\lambda, \psi^\dagger)$.  Therefore it will
make sense to restrict the theory to this subspace.

Changing variables from $M_{ij}$ to the $N$ eigenvalues
$\lambda_i$ and the $N(N-1)/2$ angular variables on which $U$ depends,
one gets the decomposition (see Dabholkar [13])
$$
  {\partial\over{\partial M_{ij}}} = \sum_m U^\dagger_{jm}U_{mi}\,
     {\partial\over{\partial \lambda_m}}
     + \sum_{k\ne l}{{U_{ki}U_{jl}^\dagger\,\tilde A_{kl}}\over
         {(\lambda_k - \lambda_l)}},    \eqno \eq
$$
where the angular derivative $\tilde A_{kl}$ is defined by $\tilde
A_{kl} \equiv \sum_m U_{lm} \,\partial/\partial U_{km}$.

In terms of the new variables the supercharges \matrixQ\ become
$$
\eqalign{
  Q        &= \sum_m \hat\Psi^*_{mm}\left(-i\,
               {\partial\over{\partial\lambda_m}} - i\,
                {{\partial \bar W(\lambda)}\over{\partial \lambda_m}}
                \right) -i \sum_{k\ne l}
                {{\hat\Psi^*_{kl}\,\tilde A^*_{kl}}\over{(\lambda_k -
                \lambda_l)}}, \cr
  Q^\dagger &= \sum_m \hat\Psi_{mm}\left(-i\,
               {\partial\over{\partial\lambda_m}} + i\,
                {{\partial \bar W(\lambda)}\over{\partial \lambda_m}}
                \right) -i \sum_{k\ne l}
                {{\hat\Psi_{kl}\,\tilde A_{kl}}\over{(\lambda_k -
                \lambda_l)}}, \cr
}  \eqn \changedQ
$$
where $\hat\Psi \equiv U \Psi U^\dagger$.  Using
$\hat\Psi_{mm} = \psi_m$, one can write a wave function depending
only on $(\lambda, \psi^\dagger)$ as
$$
  \phi (\lambda, \psi^\dagger) \equiv f(\lambda) \prod_k
       \psi^\dagger_{m_k}|0\rangle = f(\lambda) \prod_k
         \hat\Psi^*_{m_k m_k}|0\rangle.
  \eqn \WfInSubspace
$$
With a little patience one can now show that the subspace of
such
wavefunctions is indeed invariant under the action of the supercharges
\changedQ\ and that in the subspace the supercharges reduce to
$$
\eqalign{
  Q &= \sum_i \psi_i^\dagger\left( -i\,{\partial\over{\partial
\lambda_i}} - i\,
       {{\partial \bar W(\lambda)}\over{\partial\lambda_i}}\right),
\cr
  \bar Q &= \sum_i \psi_i\left( -i\,{\partial\over{\partial
\lambda_i}}
             +i\,{{\partial
             \bar W(\lambda)}\over{\partial\lambda_i}}
             - 2i\sum_{l<i}{1\over{\lambda_i - \lambda_l}}
             \right). \cr
}  \eqn \subspaceQ
$$
At first glance these expressions may seem inconsistent in that $\bar
Q$ appears to have lost its property of conjugacy to $Q$.
This paradox is resolved by the
observation
that $Q$ and $\bar Q$ are indeed hermitean conjugates with respect to
the inner
product on the original Hilbert space.  The original inner product
reduces to a nontrivial inner product on the subspace.

To see how this works, let us review the analysis of Jevicki and
Sakita [1], modified here to include fermionic degrees of freedom.
Take $\phi_1$ and $\phi_2$ to be of the form \WfInSubspace.  The inner
product on the original Hilbert space can be written as
$$
\eqalign{
  \langle \phi_1 |\phi_2\rangle
    &=  \int (d M)(d\Psi^\dagger)(d\Psi)\,
           e^{-{\rm Tr}
           (\Psi^\dagger\Psi)}
        \phi_1^*(\lambda,\psi)\,\phi_2(\lambda,\psi^\dagger)
\cr
    &=  \int (d\lambda)(d\hat\Psi^\dagger)
           (d\hat\Psi)\, e^{-{\rm Tr}
           (\hat\Psi^\dagger\hat\Psi)} J (\lambda)\,
        \phi_1^*(\lambda,\psi)\,\phi_2(\lambda,\psi^\dagger)
\cr
    &=  \int (d\lambda)(d\psi^\dagger)(d\psi)\,e^{-\psi^\dagger\psi}
           J (\lambda)\,
        \phi_1^*(\lambda,\psi)\,\phi_2(\lambda,\psi^\dagger),
\cr
}  \eqno \eq
$$
where $dM = \prod_i dM_{ii} \prod_{j>i}dM_{ij}\,d\bar M_{ij}$
integrates over the independent degrees of freedom of a hermitean
matrix only and where we have
used the fact that the fermionic
measure $\exp[-{\rm Tr}(\Psi^\dagger\Psi)]$ is invariant
under
$\Phi\rightarrow U\Phi U^\dagger$ (this measure has to be
included when one writes the fermionic part of the inner product as a
Berezinian integral [27]).

One sees that the inner product in \? differs from
the trivial inner product on the subspace in that one has to
include the measure $ J (\lambda)$. In general, this measure
is given by the integral
over the spurious degrees of freedom of the Jacobian associated to
the change of variables.

If we now rescale the wave functions as $\phi\rightarrow{
J}^{1/2}\phi$, we can use the trivial
inner product on the subspace provided we rescale the momenta as $p_i
\rightarrow
 J^{1/2}\,p_i\, J^{-1/2} = p_i + {i\over 2}\, {{\partial(\ln
J)}/{\partial\lambda_i}}$.

Applying this transformation to the supercharges \subspaceQ\ we find
$$
\eqalign{
  Q &\to \sum_i \psi_i^\dagger\left( -i\,{\partial\over{\partial
\lambda_i}}
       + \half\, i\, {{\partial\ln  J}\over{\partial\lambda_i}}
       - i\,
       {{\partial \bar W(\lambda)}\over{\partial\lambda_i}}\right),
\cr
  \bar Q &\to \sum_i \psi_i\left( -i\,{\partial\over{\partial
\lambda_i}}
          + \half\, i \,{{\partial\ln  J}\over{\partial\lambda_i}}
             +i\,{{\partial \bar
             W(\lambda)}\over{\partial\lambda_i}}
             - 2i\sum_{l<i}{1\over{\lambda_i - \lambda_l}}
             \right). \cr
}  \eqno \eq
$$

The most efficient way to solve for $ J$ is simply to use the fact
that now $\bar Q^\dagger = Q$ with respect to the trivial inner
product in the subspace. We find
$$
  {{\partial\ln  J}\over{\partial\lambda_i}}
             = 2\sum_{l<i}{1\over{\lambda_i - \lambda_l}},
  \eqno \eq
$$
and we get an effective theory with trivial inner product and
supercharges
$$
\eqalign{
  Q &= \sum_i \psi_i^\dagger\left(p_i
       - i\,
       {{\partial W
         (\lambda)}\over{\partial\lambda_i}}\right), \cr
  Q^\dagger &= \sum_i \psi_i\left(p_i
       + i\,
       {{\partial W
         (\lambda)}\over{\partial\lambda_i}}\right), \cr
}  \eqno \eq
$$
where the effective potential $W$ is given by
$$
\eqalign{
  W
              &= \bar W - \sum_{l<k}\ln(\lambda_k-\lambda_l).  \cr
}  \eqno \eq
$$

We therefore see that the supersymmetric matrix model, restricted to
a suitable supersymmetric generalization of the singlet subspace, has
the form of the super-Calogero model, thus completing the chain of
equivalences between the Calogero model, the
supersymmetrized collective field theory and the supersymmetric matrix
model.

\section {Explicit $N$-dependence}

\noindent
We now move on to a discussion  of the $N$-scaling of the various
terms in the theory.

Let us consider a general effective superpotential of the form
$$
  W = \int dx\, \bar W(x)\,\delx\varphi - {1\over 2}
     \int dx \int dy\, \ln |x-y|\,\delx\varphi\,\dely\varphi.
     \eqn \generalContW
$$
For the special case $\bar W(x) = {\omega\over 2}\,x^2$, this is just
a rewriting in terms of the continuum fields \DiscToCont\ of
the harmonic plus repulsive superpotential \CalogeroW\ of the
super-Calogero model.  In general $\bar W$ will depend on $N$ as in
\generalW.

Defining $\tilde v(x)\equiv \bar W'(x)$, the supercharges \rescaledQ\
are given by
$$
  \eqalign{
    Q &\equiv \int dx\,\psi^\dagger(x)\sqrt{\phi(x)}\,
       \biggl(
       {{p(x)}} - {i\over {2\phi}}
       \left((\delx\psi^\dagger)\psi-\psi^\dagger (\delx \psi)\right)
           \cr
      &+ i\,\tilde v(x)
       - i \pv dy\,{{\dely\varphi}\over{x-y}}
        \biggr),
    \cr
    Q^\dagger &\equiv \int dx\,\psi(x)\sqrt{\phi(x)}\,
       \biggl(
       {{p(x)}} - {i\over {2\phi}}
       \left((\delx\psi^\dagger)\psi-\psi^\dagger (\delx \psi)\right)
           \cr
      &- i\,\tilde v(x)
       + i \pv dy\,{{\dely\varphi}\over{x-y}}
        \biggr),
    \cr
  } \eqno \eq
$$
and the Hamiltonian \rescaledH\ is
$$
  \eqalign{
   H &= {1 \over {2}} \int{{dx}\over\phi}\,
        \left(\phi p - {i\over 2} \left[(\delx\psi^\dagger)\psi
              -\psi^\dagger (\delx \psi)\right]\right)^2 \cr
      &+ {{1} \over 2}\int dx\,\phi(x)
            \left(\pv dy\,{{\dely \varphi} \over
             {x-y}} - \tilde v(x) \right)^2 \cr
      &- \half\int dx\, [\psi^\dagger, \psi] \,\,{d\over dx}\left(
              \pv dy\,{{\dely\varphi}\over{x-y}} -
               \tilde v(x)\right)  \cr
     &+ \half\int dx\,\biggl[\psi^\dagger(x)\sqrt{\phi(x)},
        {d\over{dx}}\pv
        dy\,{{\psi(y)\sqrt{\phi(y)}}\over{x-y}}\biggr].  \cr
  }  \eqno \eq
 $$

This Hamiltonian and these supercharges depend implicitly on $N$
through $v$ and the constraint $\int \phi = N$.  To make the
$N$-dependence explicit, observe that for $W$ of the form \generalW\
the expressions  $\tilde v (\sqrt N x)/\sqrt N$ and $\tilde v'(\sqrt N
x)$
are independent of $N$.  We are therefore motivated to rescale $x \to
\sqrt N x$.  Then to get rid of the $N$-dependence in the constraint,
we must rescale $\phi \to \sqrt N \phi$.

The complete rescaling is given by
$x \rightarrow\sqrt{N}\,x$, $\phi\rightarrow\sqrt{N}\,\phi$,
$\varphi\rightarrow N\varphi$, $p\rightarrow p/N^{3/2}$ and
$\psi^{(\dagger)} \rightarrow\psi^{(\dagger)}/N^{1/4}$.
Defining $v(x) \equiv \tilde v(\sqrt N x)/\sqrt N$, which is
{\it independent\/} of $N$, we get for the
supercharges
$$
  \eqalign{
    Q &\equiv \int dx\,\psi^\dagger(x)\sqrt{\phi(x)}\,
       \biggl(
       {{p(x)}\over N} - {i\over {2\phi N}}
       \left((\delx\psi^\dagger)\psi-\psi^\dagger (\delx \psi)\right)
           \cr
      &+ i\,N\,v(x)
       - i N \pv dy\,{{\dely\varphi}\over{x-y}}
        \biggr),
    \cr
    Q^\dagger &\equiv \int dx\,\psi(x)\sqrt{\phi(x)}\,
       \biggl(
       {{p(x)}\over N} - {i\over {2\phi N}}
       \left((\delx\psi^\dagger)\psi-\psi^\dagger (\delx \psi)\right)
           \cr
      &- i\,N\,v(x)
       + i N \pv dy\,{{\dely\varphi}\over{x-y}}
        \biggr),
    \cr
  } \eqn \rescaledQ
$$
and for the Hamiltonian
$$
  \eqalign{
   H &= {1 \over {2N^2}} \int{{dx}\over\phi}\,
        \left(\phi p - {i\over 2} \left[(\delx\psi^\dagger)\psi
              -\psi^\dagger (\delx \psi)\right]\right)^2 \cr
      &+ {{N^2} \over 2}\int dx\,\phi(x)
            \left(\pv dy\,{{\dely \varphi} \over
             {x-y}} - v(x) \right)^2 \cr
      &- \half\int dx\, [\psi^\dagger, \psi] \,{d\over dx}\left(
              \pv dy\,{{\dely\varphi}\over{x-y}} - v(x)\right)  \cr
    &+ \half\int dx\,\biggl[\psi^\dagger(x)\sqrt{\phi(x)},
       {d\over{dx}}\pv
       dy\,{{\psi(y)\sqrt{\phi(y)}}\over{x-y}}\biggr].  \cr
  }  \eqn \rescaledH
 $$

The rescaled collective field satisfies the constraint
$$
  \int dx \,\phi (x) = 1,      \eqno \eq
$$
and we can therefore conclude that the Hamiltonian \rescaledH\ has no
remaining implicit dependence on $N$.

This concludes our general introduction to the continuum formulation
of the super-Calogero model.  In the next chapter we move on to a
discussion of the perturbation theory of this model.

\endpage

\chapter {PERTURBATION THEORY}

\noindent
In this chapter we write down the general perturbation theory of
the supersymmetric model given an arbitrary potential.  We show that
an infinite sequence of
higher order interactions is generated,
in contrast to the bosonic collective field theory, where
there are only cubic interactions.

The Hamiltonian \rescaledH\ of our theory was found to be
$$
  \eqalign{
   H &= {1 \over {2N^2}} \int{{dx}\over\phi}\,
        \left(\phi p - {i\over 2} \left[(\delx\psi^\dagger)\psi
              -\psi^\dagger (\delx \psi)\right]\right)^2 \cr
      &+ {{N^2} \over 2}\int dx\,\phi(x)
            \left(\pv dy\,{{\dely \varphi} \over
             {x-y}} - v(x) \right)^2 \cr
      &- \half\int dx\, [\psi^\dagger, \psi] \,{d\over dx}\left(
              \pv dy\,{{\dely\varphi}\over{x-y}} - v(x)\right)  \cr
    &+ \half\int dx\,\biggl[\psi^\dagger(x)\sqrt{\phi(x)},
       {d\over{dx}}\pv
       dy\,{{\psi(y)\sqrt{\phi(y)}}\over{x-y}}\biggr],  \cr
  }  \eqn \rescaledHtwo
 $$
with the constraint $\int \phi = 1$.

Let $\phi_0$ be an extremum of the bosonic potential subject to
the above constraint (we do not assume supersymmetry to be unbroken)
and expand around this vacuum configuration as
$$
  \phi = \delx\varphi = \phi_0 + {1\over N}\,\delx\eta, \qquad
  p \to Np.  \eqno \eq
$$
The factor $1/N$ makes all bosonic
propagators of order unity in position space, and absorbs the explicit
$N$-dependence into the vertices.  This is purely a matter of
convenience, and the factor $1/N$ could be left out without changing
the theory, a fact that can be seen by power counting of graphs in
position space, or most simply by writing the respective
Hamiltonians in terms of normalized creation and destruction
operators and noting that the resulting expressions are identical.

Expanding about $\phi_0$, terms linear in $\partial\eta$ cancel
and the Hamiltonian \rescaledHtwo\ becomes
$$
 \eqalign{
   H &=
         {N^2 \over 2}\int dx\, \phi_0(x)
          \left(\pv dy\,{\phi_0(y)\over x-y} - v(x)\right)^2 \cr
        &+\half \int dx\, \phi_0 p^2
         +{\pi^2\over 2} \int dx\,\phi_0(\delx\eta)^2
         -\half\int dx\pv dy\,{v(x)-v(y)\over
           x-y}\,\delx\eta\,\dely\eta   \cr
        &+ {1\over 2}\int
          dx\,\left[\psi^\dagger(x)\sqrt{\phi_0(x)},
          {d\over{dx}}\pv
          dy\,{{\psi(y)\sqrt{\phi_0(y)}}\over{x-y}}\right]\cr
     & - \half\int dx\,[\psi^\dagger(x),\psi(x)]\,{d\over dx}
          \left(\pv dy\,{\phi_0(y)\over x-y} - v(x)\right) \cr
     &+ {1\over{2N}}\int dx\,(\delx\eta)\,p^2
       + {1\over 2N}\int dx\, \delx\eta\left(\pv dy\,{\dely\eta\over
           x-y}\right)^2 \cr
     &- {i\over {2N}}\int dx\, \left[(\delx\psi^\dagger)\psi
           -\psi^\dagger (\delx \psi)\right] p   \cr
     &- {1\over{2N}}\int dx\,[\psi^\dagger(x),\psi(x)]\,
      {d\over dx} \pv dy\,{{\dely\eta}\over{x-y}}
         \cr
     &+ {1\over 2}\int
        dx\,\left[\psi^\dagger(x)\sqrt{\phi_0(x)}
        \left(\sqrt{{{\phi}/{\phi_0}}}(x)-1\right)
        ,
        {d\over dx}\pv
        dy\,{{\psi(y)\sqrt{\phi_0(y)}}\over{x-y}}\right]\cr
     &+ {1\over 2}\int
        dx\,\left[\psi^\dagger(x)\sqrt{\phi_0(x)},
        {d\over{dx}}\pv
        dy\,{{\psi(y)\sqrt{\phi_0(y)}}\over{x-y}}
        \left(\sqrt{{{\phi}/{\phi_0}}}(y)-1\right)
        \right]\cr
     &+ {1\over{8N^2}}\int{{dx}\over{\phi_0 + {1\over N}\delx\eta }}\,
        \left[\psi(\delx\psi^\dagger)
        (\delx\psi)\psi^\dagger + \psi^\dagger(\delx\psi)
        (\delx\psi^\dagger)\psi\right]  \cr
     &+ {1\over 2}\int
        dx\,\biggl[\psi^\dagger(x)\sqrt{\phi_0(x)}
        \left(\sqrt{{{\phi}/{\phi_0}}}(x)-1\right)
        ,    \cr
    &\qquad    {d\over dx}\pv
        dy\,{{\psi(y)\sqrt{\phi_0(y)}}\over{x-y}}
        \left(\sqrt{{{\phi}/{\phi_0}}}(y)-1\right)
        \biggr].\cr
 }   \eqn \expandH
$$

Some remarks are in order at this point:
In the bosonic collective field theory,
cubic terms such as appear in the bosonic part of the Hamiltonian
\rescaledHtwo\ are simplified using the identity
$$
 \int dx\,\phi(x)
   \left(\pv dy\,{\phi(y) \over {x-y}}\right)^2
 = {{\pi^2}\over 3}\int dx\,\phi^3(x),   \eqn \unregId
$$
which can readily be demonstrated [17] in Fourier space
using
$$
  \pv dy\, {{e^{iky}}\over{x - y}}
   = -\pi i \,\epsilon(k)\, e^{ikx}.  \eqno \eq
$$

In applying \unregId\ one must, however, be careful.  Obviously
as it stands it cannot be valid if the integral
$\int \phi^3$ on the right hand side diverges.
This can indeed happen, both at the level of the vacuum density
$\phi_0$,
as we shall see in the potential free case (chapter 4) and at the
level of the fluctuations $\partial\eta$, as we shall see in the
harmonic
case (chapter 5).  To be general, we therefore {\it avoid\/} using
the identity
\unregId\ in this chapter, and treat issues of regularization later,
as they arise.

On the other hand, in \expandH\ we have rewritten the quadratic terms
by applying the  identity
$$
\eqalign{
 \int dx\,\phi_0(x)
   \left(\pv dy\,{\dely\eta \over {x-y}}\right)^2
   &+ 2\int dx\,\delx\eta
       \left(\pv dy\,{\phi_0(y) \over {x-y}}\right)
       \left(\pv dz\,{\delz\eta \over {x-z}}\right)  \cr
   &=
      {\pi^2} \int dx\,\phi_0(\delx\eta)^2,  \cr
} \eqno \eq
$$
which is easy to show in Fourier space.
For all the cases that we will consider, the integrals in \? are
well behaved and no special regularizations are needed.

The first term in the Hamiltonian \expandH\ is just a constant.  If it
is nonzero,
then supersymmetry is broken to leading order in $N$.  Conversely, if
the vacuum configuration of the density field satisfies
$$
  V_{\rm eff}(\phi_0) = {{N^2} \over 2}\int dx\,\phi_0(x)
     \left(\pv dy\,{{\phi_0(y)} \over {x-y}} - v(x)\right)^2 = 0.
  \eqno \eq
$$
then supersymmetry is preserved to leading order.
This will be the case if
$$
         \pv dy\,{{\phi_0(y)} \over
             {x-y}} - v(x) = 0.
  \eqn \nonFreeCond
$$

Restricting attention to the bosonic piece, in addition to the terms
associated in the bosonic string theory [2] to a massless
scalar particle, one has the additional contribution
$$
         -\half\int dx\pv dy\,{v(x)-v(y)\over
           x-y}\,\delx\eta\,\dely\eta,
$$
which may in general affect
the dynamics in unexpected ways.  However, using the fact that $\int
dx\, \delx\eta = 0$, it is easy to show that for potentials of the
general form $v(x) = a + bx + cx^2$, this term falls away and the
quadratic spectrum is that of a massless scalar. These potentials
include the free case, the harmonic
case, and the Marinari-Parisi potential [13, 32].

Using
$$
  \sqrt{{{\phi}/{\phi_0}}}-1 = {1\over{2N\phi_0}}\,\delx\eta
    -{1\over{8N^2\phi_0^2}}\, (\delx\eta)^2 + o\left({1\over
    {N^2}}\right),     \eqno \eq
$$
the Hamiltonian can be written up to cubic order as
$$
 \eqalign{
   H &=  {N^2 \over 2}\int dx\, \phi_0(x)
          \left(\pv dy\,{\phi_0(y)\over x-y} - v(x)\right)^2 \cr
     &+ {1\over 2} \int dx\,\phi_0 p^2 + {{\pi^2}\over 2}\int dx\,
        \phi_0(\delx\eta)^2
         -\half\int dx\pv dy\,{v(x)-v(y)\over
           x-y}\,\delx\eta\,\dely\eta   \cr
    &+ {1\over 2}\int
          dx\,\left[\psi^\dagger(x)\sqrt{\phi_0(x)},
          {d\over{dx}}\pv
          dy\,{{\psi(y)\sqrt{\phi_0(y)}}\over{x-y}}\right]  \cr
     & - \half\int dx\,[\psi^\dagger(x),\psi(x)]\,{d\over dx}
          \left(\pv dy\,{\phi_0(y)\over x-y} - v(x)\right) \cr
     &+ {1\over{2N}}\int dx\,(\delx\eta)\,p^2
        + {{\pi^2}\over{6N}}\int dx\,(\delx\eta)^3
        - {i\over {2N}}\int dx\, \left[(\delx\psi^\dagger)\psi
           -\psi^\dagger (\delx \psi)\right] p \cr
     &- {1\over{2N}}\int dx\,[\psi^\dagger(x),\psi(x)]\,
         {d\over{dx}}\pv dy\,{{\dely\eta}\over{x-y}}
           \cr
     &+ {1\over{4N}}\int
        dx\,\left[{\psi^\dagger(x)\over\sqrt{\phi_0(x)}}\,
        \delx\eta
        ,
        {d\over{dx}}\pv
        dy\,{{\psi(y)\sqrt{\phi_0(y)}}\over{x-y}}\right]  \cr
     &+ {1\over{4N}}\int
        dx\,\left[{\psi^\dagger(x)\sqrt{\phi_0(x)}},
        {d\over{dx}}\pv
        dy\,{{\psi(y)/\sqrt{\phi_0(y)}}\over{x-y}}\,
        \dely\eta
        \right]
+ o\left({1\over N^2}\right),
\cr
 }   \eqn \perturbH
$$
and the supercharges  \rescaledQ\ can be written as
$$
 \eqalign{
   Q &= \int
        dx\,\sqrt{\phi_0}\,\psi^\dagger(x)\sqrt{{\phi}/{\phi_0}}\,
        \biggl\{
        p(x) - i\pv dy\,{{\dely\eta}\over{x-y}} \cr
      &\quad\, -i N \left(\pv dy\,{\phi_0(y)\over x-y} - v(x)\right)
\cr
      &\quad\,
        - {i\over{2N\phi_0(1+{1\over{N\phi_0}}\delx\eta)}}
        \left((\delx\psi^\dagger)\psi - \psi^\dagger(\delx\psi)\right)
        \biggr\}                         \cr
     &= -iN \int dx\,\sqrt{\phi_0}\, \psi^\dagger
          \left(\pv dy\,{\phi_0(y)\over x-y} - v(x)\right)
           \cr
       &\quad\, + \int dx\,\sqrt{\phi_0}\, \psi^\dagger(x)
        \biggl\{
        p(x) - i\pv dy\,{{\dely\eta}\over{x-y}}  \cr
       &\quad\,      - {i\over 2\phi_0}\,\delx\eta
                \left(\pv dy\,{\phi_0(y)\over x-y} - v(x)\right)
          \biggr\} \cr
     &\quad\, + {1\over{2N}}\int
        {dx\over\sqrt{\phi_0}}\,\psi^\dagger(x)
        \biggl\{\delx\eta\biggl[
        p(x) - i\pv dy\,{{\dely\eta}\over{x-y}}   \cr
     &\quad\, +
           {i\over 4\phi_0}\,\delx\eta
           \left(\pv dy\,{\phi_0(y)\over x-y} - v(x)\right)
        \biggr] \cr
     &\quad\, - i
        \left((\delx\psi^\dagger)\psi - \psi^\dagger(\delx\psi)\right)
        \biggr\} + o\left({1\over N^2}\right),             \cr
  Q^\dagger &= {\rm h.c.}. \cr
 }  \eqn \perturbQ
$$

We see that the system develops an infinite sequence of polynomial
interactions in the bare string coupling constant $1/N^2$.  This is in
contrast to the cubic bosonic collective field theory Hamiltonian [2,
6, 7] and
does not depend on the presence of a potential $v$.  The supercharges
also acquire expansions to all orders in perturbation theory, typical
of supersymmetric theories expanded about background configurations.
In general, we expect the presence of a nontrivial background
$\phi_0$ to \lq\lq
dress\rq\rq\ the coupling constant and allow us to take a suitable
double scaling limit.

One of the remarkable properties of the
cubic bosonic Hamiltonian is that [7] the integral representation
of a given amplitude can in general be reinterpreted as a sum of
standard tachyon interchange diagrams, plus contact terms, in a one to
one correspondence to first quantized Liouville computations [9].
In studies of critical closed string field theory, an infinite
sequence of polynomial interactions seems to be required to obtain
agreement with the first quantized integrations over moduli space
[36].  In the model discussed in this thesis, the need to also
include an
infinite sequence of vertices is unavoidable.  It should
be clear that the reason for the supersymmetrized version of the
simple bosonic cubic Hamiltonian to contain an infinite sequence of
higher order vertices with derivative couplings is that the
supercharges themselves have an infinite perturbative expansion.

It is conceivable that other supersymmetric extensions of the
bosonic collective
string field Hamiltonian (possibly formulated directly in the
continuum) may exist, with properties different from those described
here.  Ultimately, once a genuine field theory of $d=1$ superstrings
is formulated, the correct choice would be selected by requiring
agreement with the super-Liouville theory [15].

After an analysis of the free and the harmonic theories in
chapters 4 and 5, we shall return to the general potential case
in chapter 6.

\endpage

\chapter {THE POTENTIAL FREE CASE}

\section {Motivation}

\noindent
In this chapter we specialize the
discussion to the potential free case characterised by $v(x) = 0$.

One reason for studying this case is the following:  we saw in the
previous chapter
that any nontrivial potential generates a nontrivial perturbative
background.
In the bosonic case it is well known that the inverted harmonic
oscillator potential is associated with $d=1$ strings.
By starting with the theory without
a potential term, however, in the bosonic case one can obtain an
inverted
harmonic oscillator potential by means of a simultaneous field
transformation and coordinate reparametrization [16].
The potential free case therefore
corresponds to a background independent formulation of the theory.

Also, the bosonic potential free model has been solved
exactly [16] and stringy semiclassical solutions have been described.
Some of these results turn out to be easily generalizable to the
supersymmetric
case, as we now demonstrate.

\section {The vacuum density}

\noindent
As a first step, we discuss the extremization of the potential.  In
the process we shall see that to make the free theory well defined,
one needs to introduce a regularization, a fact that was already
alluded to in the previous chapter.

The potential term of the Hamitonian \rescaledHtwo\ is given in the
free case by
$$
  V_{\rm eff}(\phi) = {{N^2} \over 2}\int dx\,\phi(x)
     \left(\pv dy\,{{\phi(y)} \over {x-y}}\right)^2,
  \eqn \freeV
$$
As we noted in the previous chapter,
for supersymmetry to be unbroken to leading order in $N$, the
vacuum density $\phi_0$ must satisfy
$$
  V_{\rm eff} (\phi_0) = 0.  \eqno\eq
$$

Alternatively, one may try to obtain $\phi_0$ by applying the identity
\unregId\ to rewrite the potential \freeV\ as a local expression
in $\phi$ and then
extremizing with respect to $\phi$.  However, in the free case the
integral $\int \phi^3$ will in general diverge and has to be
regulated.  To see why,
note that
on physical grounds one expects the density of eigenvalues
to be constant,
in which case equation \? follows from
the result
$$
  \pv {dy \over{x-y}} = 0.  \eqn \freeCond
$$
For a constant $\phi_0$ the integral $\int\phi_0^3$ does not converge
and the identity $\unregId$ becomes invalid.
In addition to this, the vacuum
density cannot consistently be normalized so as to satisfy the
constraint $\int\phi_0 = 1$.

To overcome these problems, we can regulate the system by putting it
in
a box of length $L$ and imposing periodic boundary conditions.  More
precisely, we restrict the original theory to the subspace of all
states $|\Psi\rangle$ such that $\phi(x)|\Psi\rangle =
\phi(x+L)|\Psi\rangle$.
For this restriction to be consistent, the periodic subspace must be
invariant under the supercharges.  To see that this is indeed the
case, note that for states $|\Psi\rangle$ satisfying the above
periodicity condition we have
$$
\eqalign{
  \pv_{\!\!\!\!-\infty}^\infty dy\,{{\phi(y)}\over{x-y}} |\Psi\rangle
        &= \sum_n \pv_{\!\!\!\! -L/2}^{L/2}
            dy\,{{\phi(y)}\over{x-y+nL}}|\Psi\rangle \cr
        &= \pv_{\!\!\!\! -L/2}^{L/2} dy\,\phi(y) \sum_n
            {1\over{x-y+nL}}|\Psi\rangle\cr
        &= \pv_{\!\!\!\! -L/2}^{L/2} dy\,\phi(y)\, {\pi\over L}\,
           \cot \biggl[ {{(x-y)\pi}\over L} \biggr]|\Psi\rangle,
}  \eqno \eq
$$
where we have used the identity $\sum_n {1\over{x+n\pi}} = \cot x$,
which is almost trivial to see by comparing pole structures.

Using \?, the supercharge $Q$ in \rescaledQ\ becomes, on the periodic
subspace,
$$
  \eqalign{
    Q &\equiv \int_{-\infty}^\infty
            dx\,\psi^\dagger\sqrt{\phi}\,
       \biggl(
       {{\sigma}\over N}
       + i\,N\,v
       -i\,N \pv_{\!\!\!\! -L/2}^{L/2} dy\,\phi(y)\, {\pi\over L}\,
                \cot \left[ {{(x-y)\pi}\over L} \right]
        \biggr)
    \cr
    &= \int_{-L/2}^{L/2}
            dx\,\psi^\dagger\sqrt{\phi}\,
       \biggl(
       {{\sigma}\over N}
       + i\,N\,v
       -i\,N \pv_{\!\!\!\! -L/2}^{L/2} dy\,\phi(y)\, {\pi\over L}\,
                \cot \left[ {{(x-y)\pi}\over L} \right]
        \biggr)
    \cr
  }   \eqno \eq
$$
up to an infinite constant, provided the potential $v$ has been chosen
to be
periodic. Due to the fact that $\cot{(\pi x/L)}$  has period $L$, we
see that the periodic subspace is indeed invariant under $Q$, and that
we can restrict to the interval $[-L/2,L/2]$ provided we change the
kernel
$$
  {1\over{x-y}} \rightarrow {\pi\over L}\cot \left[ {{(x-y)\pi}\over
   L} \right].   \eqno \eq
$$

Not suprisingly, this kernel is the one that appears in quantized
matrix models of unitary matrices.  The reason for this is that when
one restricts the original theory with (Fourier space) collective
variables $\phi_k = \Tr \left(e^{ikM}\right)$ to the periodic
subspace, the collective variables in the subspace are those of a
theory of unitary matrices: they are given by $\phi_n = \Tr
\left(U^n\right)$, where
the unitary matrix $U \equiv e^{2\pi iM/L}$.

Henceforth we shall interpret the kernel
$$
  (K\phi)(x) = \pv dy\,{{\phi(y)}\over{x-y}}
$$
as the limit as $L \rightarrow\infty$ of
$$
  (K'\phi)(x) = \pv dy\,\phi(y){\pi\over L}
                \cot \biggl[ {{(x-y)\pi}\over L} \biggr].  \eqn
\boxKernel
$$

In this sense, the equation \freeCond\ remains true in a box of
finite length $L$, and the identity that should replace \unregId\ is
given by
$$
 \int dx\,\phi(x)
   \left(\pv dy\,{\phi(y) \over {x-y}}\right)^2
 = {{\pi^2}\over 3}\int dx\,\phi^3(x)
    - {{\pi^2}\over{3L^2}}\left( \int dx\,\phi(x) \right)^3.
   \eqn \regId
$$
This can be shown in Fourier space, the second term arising from a
careful treatment of the zero mode.  The same result was proven by
analytical methods in [19].
For a density $\phi$ that is normalisable independently of $L$, the
second term vanishes in the limit $L \to\infty$, which is consistent
with the original identity \unregId.  However, in the example of the
free case the
vacuum density is constant and should be normalised as $\phi_0 = 1/L$,
and we see that both terms on the right hand side of \regId\ are of
order $1/L^2$.

The second term on the right hand side of \regId\ is commonly
dropped in the bosonic $d=1$ string, as it only shifts the value of
the energy by a constant value (the energy of an ensemble of
$\int dx\,\phi(x)$ free fermions in a box of length $L$).
In the supersymmetric formalism, however, we cannot
consistently leave it out.

Using the identity \regId, we find
$$
  V_{\rm eff}(\phi)
   =  {{N^2\pi^2}\over 6}\int dx\,\phi^3(x)
      - {{N^2\pi^2}\over{6L^2}}\left( \int dx\,\phi(x) \right)^3.
           \eqno\eq
$$
The background $\phi_0$ satisfies the stationarity condition
$$
  {{\pi^2\phi_0^2}\over 2} - {{\pi^2}\over{2L^2}}
  \left( \int dx\,\phi_0(x) \right)^2 = 0.
  \eqn \stationarity
$$
Requiring $\int\phi_0 = 1$, we find  $\phi_0=1/L$.

Comparing this with the bosonic case, where
a constant background $\phi_0$ results from an
effective potential
$$
  V_B(\phi) = N^2 \left[{{\pi^2}\over 6} \int dx\,\phi^3(x)
              - \mu_F\int dx\,\phi(x)
              \right]
  \eqno \eq
$$
for which the stationarity condition yields
$$
 {{\pi^2}\over 2}\phi^2 - \mu_F = 0, \eqno \eq
$$
one sees that in equation \stationarity\ one can identify
$$
  \mu_F \equiv {{\pi^2}\over{2L^2}}
              \left( \int dx\,\phi(x) \right)^2.
$$

This $\mu_F$ is the fermi energy one would expect for
a system
of $\int dx\,\phi(x)$ one dimensional free particles.
In this sense the chemical
potential can be said to be \lq\lq dynamically induced\rq\rq\ in the
supersymmetric system.

Using $\phi_0 = 1/L$, we immediately find that $V_{\rm eff}(\phi_0) =
0$, so that supersymmetry is unbroken to leading order.

To summarize, due to the fact that in the free case the potential is
unbounded,
we needed
to introduce a periodic regularization
to make the theory well-defined.  Such a regularization changes
the identity \unregId\ to give \regId.

Having found the vacuum density, we can now expand about it to find
the effective perturbation theory for the fluctuations, which will be
done in the next section.

\section {Perturbative expansion}

\noindent
We are now ready to specialize the general perturbative expansion of
the
previous chapter to the free case.

Applying the identity \regId\ to the fluctuations $\partial\eta$ and
using the fact that, due to the constraint $\int\partial\eta = 0$,
the Hamiltonian \perturbH\  becomes
$$
 \eqalign{
   H &= {{\phi_0}\over 2} \int dx\, p^2 + {{\pi^2}\over 2}\int dx\,
        (\delx\eta)^2
        + {{\phi_0}\over 2}\int
          dx\,\left[\psi^\dagger(x),
          {d\over{dx}}\pv
          dy\,{{\psi(y)}\over{x-y}}\right]\cr
     &+ {1\over{2N}}\int dx\,(\delx\eta)\,p^2
        + {{\pi^2}\over{6N}}\int dx\,(\delx\eta)^3
        - {i\over {2N}}\int dx\, \left[(\delx\psi^\dagger)\psi
           -\psi^\dagger (\delx \psi)\right] p \cr
     &- {1\over{2N}}\int dx\,\left[\psi^\dagger(x),\psi(x)
         {d\over{dx}}\pv dy\,{{\dely\eta}\over{x-y}}\right] \cr
     &+ {1\over{4N}}\int
        dx\,\left[\psi^\dagger(x)\,
        \delx\eta
        ,
        {d\over{dx}}\pv
        dy\,{{\psi(y)}\over{x-y}}\right]\cr
     &+ {1\over{4N}}\int
        dx\,\left[\psi^\dagger(x),
        {d\over{dx}}\pv
        dy\,{{\psi(y)}\over{x-y}}\,
        \dely\eta
        \right]
+ o \left( {1\over N^2} \right),
\cr
}   \eqn \freeH
$$
and the supercharges \perturbQ\ are
$$
 \eqalign{
   Q
     &= \sqrt{\phi_0}\int
        dx\,\psi^\dagger(x)
        \left[
        p(x) - i\pv dy\,{{\dely\eta}\over{x-y}} \right] \cr
     &\quad + {1\over{2N\sqrt{\phi_0}}}\int
        dx\,\psi^\dagger(x)
        \biggl[\delx\eta\left(
        p(x) - i\pv dy\,{{\dely\eta}\over{x-y}}\right) \cr
     &\quad - i
        \left((\delx\psi^\dagger)\psi - \psi^\dagger(\delx\psi)\right)
        \biggr]
+ o \left({1\over N^2} \right),
\cr
  Q^\dagger &= {\rm h.c.} \cr
 }  \eqno \eq
$$

The double scaling limit can now be taken by letting $N\to\infty$ and
$L\to\infty$ while keeping the dressed string coupling constant
$g_{\rm st} = 1/(N\phi_0^2)$ constant.
This is best seen by
defining $d\tau = dx/\phi_0$ and rescaling $p\to p/\phi_0$, $\psi \to
\psi/\sqrt{\phi_0}$, $\psi^\dagger\to\psi^\dagger/\sqrt{\phi_0}$.
The length of the \lq\lq extra
dimension\rq\rq\ (\ie, the range of integration of $\tau \equiv
x/\phi_0 = Lx$) becomes infinite in the scaling limit.

We now proceed to obtain the spectrum as well as the three string
vertices in a creation-annihilation basis.  For this we introduce the
expansions
$$
\eqalign{
\eta(x,t) &= {1\over{2\pi}}\sum_{n\ne 0}{1\over{\sqrt{|n|}}}
  \left(
   a_n(t)\, e^{{2\pi inx}/ L} + a_n^\dagger(t)\, e^{-{{2\pi inx}/
       L}}
  \right),   \cr
p(x,t) &= {{2\pi}\over L} \sum_{n\ne 0}{{\sqrt{|n|}}\over{2i}}
  \left(
   a_n(t)\, e^{{2\pi inx}/ L} - a_n^\dagger(t)\, e^{-{{2\pi inx}/
       L}}
  \right),  \cr
} \eqno \eq
$$
$$
\psi(x) = {1\over{\sqrt L}}\sum_{n\ne 0}b_n(t)\,e^{{2\pi inx}/ L},
\quad
\psi^\dagger(x) = {1\over{\sqrt L}}\sum_{n\ne
        0}b_n^\dagger(t)\,e^{-{{2\pi inx}/ L}}
\eqno \eq
$$
with (anti-)commutation relations
$$
 [a_n,a_m^\dagger] = \delta_{mn}, \qquad
 [a_n,a_m] = [a_n^\dagger,a_m^\dagger] = 0,  \eqno\eq
$$
$$
 \{ b_n,b_m^\dagger \} = \delta_{mn}, \qquad
 \{ b_n,b_m \} = \{ b_n^\dagger,b_m^\dagger \} = 0.
 \eqno \eq
$$

One then obtains, after some algebra
$$
  H = H_0{}^B + H_0{}^F + {1\over N}H_3{}^B + {1\over N}H_3{}^F +
       o\left({1\over{N^2}}\right),
$$
with
$$
  H_0{}^B = {{2\pi^2}\over L}\phi_0\sum_n|n|\,(a_n^\dagger a_n+\half),
$$
$$
  H_0{}^F = {{2\pi^2}\over L}\phi_0\sum_n|n|\,(b_n^\dagger b_n-\half),
$$
$$
  H_3{}^B = {i\over 2}\left({{2\pi}\over L}\right)^2
  \sum_{{n_1,n_2>0} \atop {n_1,n_2<0}}
   \sqrt{n_1n_2|n_1+n_2|}\,\epsilon(n_1+n_2)\,
   (a_{n_1+n_2}^\dagger a_{n_1} a_{n_2} - {\rm h.c.}),
$$
$$
 \eqalign{
  H_3{}^F &= i\left({{2\pi}\over L}\right)^2
             \sum_{{n_1,n_2>0} \atop {n_1,n_2<0}}
             n_1\sqrt{|n_2|}\,
             (b^\dagger_{n_1+n_2}b_{n_1}a_{n_2} - b_{n_1}^\dagger
             b_{n_1+n_2}a_{n_2}^\dagger)   \cr
          &- {i\over 2}\left({{2\pi}\over L}\right)^2
             \sum_{{n_1,n_2>0} \atop {n_1,n_2<0}}
             n_2\sqrt{|n_2|}\,
             (b^\dagger_{n_1}b_{n_1+n_2}a_{-n_2} -
             b_{n_1+n_2}^\dagger
             b_{n_1}a_{-n_2}^\dagger)   \cr
          &- {i\over 2}\left({{2\pi}\over L}\right)^2
             \sum_{{n_1,n_2>0} \atop {n_1,n_2<0}}
             n_1\sqrt{|n_1+n_2|}\,
             (b^\dagger_{n_2}b_{-n_1}a_{n_1+n_2} - b_{-n_1}^\dagger
             b_{n_2}a_{n_1+n_2}^\dagger).   \cr
 }  \eqn \oscillatorH
$$

The presence of Majorana fermions is evident from the quadratic part
of equation \oscillatorH.

To summarise, in this section we discussed the perturbation
theory for the free case.  We exhibited the explicit two and three
point functions and identified the fluctuations to be at the
semiclassical level those of a massless scalar and a massless Majorana
fermion.

\section {Comparison with supergravity}

\noindent
It is instructive to try to understand the supersymmetric collective
field theory with reference to two dimensional supergravity and
non-linear sigma models.

Neglecting the Lagrange multiplier contribution, we consider the
bosonic collective string field Lagrangian
$$
  {{L_B}\over{N^2}} = \half\int{{dx}\over\phi}\,{\dot\varphi}^2
     - {{\pi^2}\over 6}\int dx\,(\delx\varphi)^3.
  \eqn \bosonicL
$$
Defining a two dimensional metric
$$
  g_{\alpha\beta} (\phi) \equiv
  \pmatrix{
        \delx\varphi & 0 \cr
        0 & -{3\over{\,\pi^2}}({1\over{\delx\varphi}}) \cr
  },
  \eqn \bosonicG
$$
we notice that the action associated with the Lagrangian \bosonicL\
can be written as
$$
 {{S_B}\over{N^2}} = \half\int dx\int dt\,g^{\alpha\beta}(\phi)\,
   \partial_\alpha\varphi\,\partial_\beta\varphi.
  \eqn \bosonicS
$$

It has already been pointed out elsewhere [2, 6] that, in the presence
of
a potential, the quadratic action is that of a scalar tachyonic field
coupled to a nontrivial classical background, and that the one
loop quantum correction to the ground state energy can be understood
entirely as resulting from an anomaly in the reparametrization that
transforms
the metric to a flat one.
It turns out that in the potential free case, the
full action is of this form.

Because in two dimensions both the target space and the world sheet
have the same dimension, there are two ways in which one can look at
the action \bosonicS.  One is to regard it as a nonlinear sigma model
for
the centre of mass coordinate (tachyon) of the string.  In this case,
the Lagrange multiplier contribution can be thought of as being of a
topological nature, since $\int dx\, \delx\varphi = 1$.  The ${1
/N}$ expansion then corresponds to an expansion about a
translational noninvariant background.  Physically, $\phi_0 = \text
{constant}$ implies that $\varphi_0 \sim x$, and interchanging the
roles of $x$ and $t$ (it is well known that the collective space-time
coordinates are interchanged in comparison with the first quantized
Liouville approach, for instance), one has $\varphi_0 \sim t$ plus
oscillator contributions.  In this picture, the zero mode can be
associated with a classical uniform motion of the centre of mass.

Alternatively, since $\det g = {\rm constant}$, one can think of
\bosonicS\ as an action for the matter field $\varphi$ in a gauge
fixed
gravitational background.  It would be of interest, of course, to
establish whether the residual simmetries of \bosonicS\ are related to
the symmetry algebra uncovered in [4, 5] and also whether the
identification \bosonicG\ is consistent with recently proposed
effective actions for the gravitational states of the theory [24].
These questions, however, are outside the scope of this thesis.

For our purposes, given that there is a well defined procedure that
generalizes the action of a two dimensional scalar field in a
gravitational background to two dimensional supergravity, we ask
ourselves if the supersymmetric collective field theory corresponds to
two dimensional supergravity in a gauge consistent with \bosonicG.

Before doing so, we introduce some notation.  We notice that
$$
  {i\over \pi}\,{d\over{dx}}\pv dy\,{{e^{iky}}\over{x-y}}
    = \epsilon (k)\, {d\over{dx}}\,e^{ikx} = i|k|e^{ikx}.
  \eqno \eq
$$
It is therefore natural to define the operator
$$
\adelx\xi \equiv {i\over\pi}{d\over{dx}}\pv dy\,
  {{\xi(y)}\over{x-y}},
\eqno \eq
$$
which we will call the \lq\lq absolute\rq\rq\ derivative.  Its
properties include
$$
\adelx e^{ikx} = i|k|e^{ikx},\qquad \adelx e^{-ikx} = i|k|e^{-ikx},
$$
$$
\adelx \xi^\dagger = -(\adelx\xi)^\dagger, \qquad
\pv dy\,{{\dely\xi}\over{x-y}}=-i\pi\adelx\xi. \eqno \eq
$$
Suggestive as the notation may be, it should be noted that the
absolute derivative does not obey a Leibnitz rule.

We rewrite the action corresponding to the Lagrangian \symmetricL\ in
the free case in terms of the absolute derivative:
$$
 \eqalign{
  S &= \half\int dx\int dt\,g^{\alpha\beta}(\phi)\,
         \partial_\alpha\varphi\,\partial_\beta\varphi \cr
     &+ {i \over 2} \int{{dx} \over
          \phi}\,(\psi^\dagger\delt\psi
             - {\delt \psi}^\dagger \psi)  \cr
     &+ {i \over 2} \int{{dx} \over {\phi^2}}\,\delt\varphi\,
         (\delx\psi^\dagger\psi
          -\psi^\dagger \delx \psi)  \cr
     &+ i\pi\int dt\int dx\,\psi^\dagger\adelx\psi
      - i\pi\int dt\int{{dx}\over{\phi}}\,\psi^\dagger\psi\,
             \adelx\phi.   \cr
 }
  \eqno \eq
$$
In terms of real Majorana components ($\psi={1\over{\sqrt 2}}(\psi_1 -
i\psi_2 )$),
$$
 \eqalign{
  S &= \half\int dx\int dt\,g^{\alpha\beta}(\phi)\,
         \partial_\alpha\varphi\,\partial_\beta\varphi \cr
     &+ {i \over 2} \int\,{{dx} \over \phi}\,(\psi_1\delt\psi_1
             + \psi_2\delt\psi_2)  \cr
     &- {i \over 2} \int\,{{dx} \over {\phi^2}}\,\delt\varphi\,
         (\psi_1\delx\psi_1
          +\psi_2\delx \psi_2) \cr
     &- {\pi\over 2}\int dt\int dx\,
         \left(\psi_2\adelx\psi_1 - \psi_1\adelx\psi_2\right) \cr
     &+ {\pi\over 2}\int dt \int{{dx}\over{\phi}}\,
           \left(\psi_2\psi_1 - \psi_1\psi_2\right)
             \adelx\phi.   \cr
 }
  \eqn \majoranaS
$$

The two dimensional supergravity action is given by
$$
\eqalign{
 S_{\rm sg} &= \half\int d^2\sigma\, e\,g^{\alpha\beta}\,
          {\partial_\alpha}\varphi\,{\partial_\beta}\varphi
         + {i\over 2}\int d^2\sigma\,e\,\bar\lambda\,\rho^\alpha
           \partial_\alpha \lambda    \cr
        &- \int d^2\sigma\, e\,\bar\chi_\alpha\,\rho^\beta\rho^\alpha
            \lambda\,\partial_\beta\varphi
         - {1\over 4}\int d^2\sigma \,e\, \bar\lambda\lambda\,
            \bar\chi_\alpha\,\rho^\beta\rho^\alpha\chi_\beta, \cr
}  \eqn \supergravityS
$$
using the conventions of [25].

The simplest gauge fixing condition consistent with \bosonicG\ is
$$
  e_\alpha^a \equiv
  \pmatrix{
        \sqrt\phi & 0 \cr
        0 & {1\over{\pi\sqrt{\phi/ 3}}}  \cr
  }.
  \eqno \eq
$$
Then the quadratic piece and the interaction terms containing time
derivatives in \supergravityS\ are given by
$$
\eqalign{
 S_{\rm sg} &= \half\int d^2\sigma\, e\,g^{\alpha\beta}\,
          {\partial_\alpha}\varphi\,{\partial_\beta}\varphi
         + {i\over 2}\int
            {{d^2\sigma\,e}\over{\sqrt{\phi}}}\,
           \left(\lambda_-\delt\lambda_-
                +\lambda_+\delt\lambda_+\right) \cr
        &+ {{i\pi}\over {2\sqrt 3}}\int
            d^2 \sigma \sqrt\phi
           \left(\lambda_-\delx\lambda_-
                -\lambda_+\delx\lambda_+\right) \cr
        &- \int{{d^2\sigma\,e}\over{\sqrt{\phi}}}\,
             (\chi_0^T\gamma^0\lambda)\,\dot\varphi
         - {\pi\over 3} \int d^2\sigma\, e\,
             (\chi_1^T\gamma^1\lambda)\,\dot\varphi + \cdots. \cr
}  \eqno \eq
$$
With a further choice $\chi_0 = i\gamma^0\delx \lambda/\phi$; $\chi_1
= 0$, the fermionic terms involving time derivatives in \? can be
brought into a form similar to those of \majoranaS\ if we let $\lambda
= {\psi/{\phi^{1/4}}}$.

One encounters several difficulties when trying to push the analogy
any further.  First, equation \supergravityS\ contains terms quartic
in the fermion fields,
which are absent from \majoranaS.
Second, in trying to match the terms involving spatial
derivatives one finds that, apart from different numerical constants,
equation \majoranaS\ contains \lq\lq absolute\rq\rq\ spatial
derivatives
which, as pointed out previously, do not obey a Leibnitz rule.

One may ask if it is possible to redefine fields so that \lq\lq
absolute\rq\rq\
derivatives are replaced by standard derivatives, especially in
view of the fact that the identity \regId\ can be stated as
$$
 \int\,dx\,\delx\varphi\,
   (\adelx\varphi)^2
 = {1\over 3}\int\,dx\,(\delx\varphi)^3
    - {1\over{3L^2}}\left( \int\,dx\,\delx\varphi \right)^3.
   \eqno \eq
$$
One finds that for a term such as
$$
-i\int dx\,\psi^\dagger\adelx\psi = {1\over 2}\int dx\,
     (\psi_2\adelx\psi_1 - \psi_1\adelx\psi_2),
  \eqno \eq
$$
such a transformation exists:  If
$$
\psi = {1\over{\sqrt 2}} \sum_n b_n e^{{{2\pi inx}/ L}}, \qquad
\psi^\dagger = {1\over{\sqrt 2}} \sum_n b_n^\dagger e^{-{{2\pi
    inx}/ L}},
\eqno \eq
$$
and we let
$$
 \eqalign{
  \psi_+ &= \sum_n {{\theta(n)}\over{\sqrt L}}\,
       \left[
         \left({{1-i}\over{\sqrt 2}}\right)b_n e^{inx}
       + \left({{1+i}\over{\sqrt 2}}\right)b_n^\dagger e^{-inx}
       \right]  \cr
  \psi_- &= \sum_n {{\theta(-n)}\over{\sqrt L}} \,
       \left[
         \left({{1-i}\over{\sqrt 2}}\right)b_n e^{inx}
       + \left({{1+i}\over{\sqrt 2}}\right)b_n^\dagger e^{-inx}
       \right],  \cr
 }    \eqno \eq
$$
then
$$
 \eqalign{
   -i\int dx\,\psi^\dagger\adelx\psi &= {1\over 2}\int dx\,
         (\psi_2\adelx\psi_1 - \psi_1\adelx\psi_2)   \cr
  &= {i\over 2}\int dx\,(\psi_-\delx\psi_- - \psi_+\delx\psi_+).
 }
  \eqno \eq
$$
However, such a procedure does not seem to be generalizable to the
interaction terms.

In conclusion, the supersymmetric collective field theory, which
provides a continuum description of the supersymmetric extension of
matrix model eigenvalue dynamics, is not reducible to a gauge fixed
two dimensional supergravity theory, unlike its bosonic counterpart.

\endpage

\section {Exact nonperturbative results}

\noindent
The bosonic part of our Hamiltonian \freeH\
coincides with the free bosonic collective field Hamiltonian,
for which a nonperturbative analysis was carried out
in [17].  There the exact, nonperturbative spectrum was obtained by
explicitly diagonalising the full bosonic Hamiltonian
 $H^B = H_0{}^B + {1\over N}H_3{}^B$.  As $H_3$ commutes with $H_0$,
the energy eigenvalues are simply the eigenvalues of the quadratic
piece plus a correction of order $1/N$.
The exact spectrum is found by exploiting the
correspondence between the collective field Hamiltonian and the matrix
model Laplacian, whose eigenstates are just the character polynomials
of $U(N)$.  A subset of these character polynomials consists of the
\lq\lq
single particle\rq\rq\ branches which can be written in the continuum
limit as
$$
\eqalign{
   |\Psi(k)\rangle &= \left(i\,\epsilon(k)\,
                   a^\dagger_k - \half\int_0^k dk'\,
                     \epsilon(k')\,\epsilon(k-k')\,
                a_{k'}^\dagger a_{k-k'}^\dagger +
                \cdots\right)|0\rangle,\cr
   |\tilde\Psi(k)\rangle &= \left(i\,\epsilon(k)\,
                   a^\dagger_k + \half\int_0^k dk'\,
                     \epsilon(k')\,\epsilon(k-k')\,
                a_{k'}^\dagger a_{k-k'}^\dagger +
                 \cdots\right)|0\rangle,\cr }   \eqn \exactStates
$$
with energies
$$
\eqalign{
  \omega(k) &= \pi\phi_0|k| + {k^2\over {2N}},   \cr
  \tilde\omega(k) &= \pi\phi_0|k| - {k^2\over {2N}},
       \quad (0 < |k| < k_F \equiv \pi\phi_0).  \cr
}   \eqn \exactEnergies
$$

Note that our expressions for the states differ from those in [17],
as there the fundamental fields for the the expansions were taken to
be
$\phi \equiv \delx\varphi$ and $\Pi \equiv - \delx p$.  Transforming
$a_k \rightarrow - i\,\epsilon(k)\,a_k$;
$a_k^\dagger \rightarrow i\,\epsilon(k)\,a_k^\dagger$,
we regain the expressions of reference [17].

The states \exactStates, being eigenstates of the
purely bosonic part of the supersymmetric Hamiltonian, are also
eigenstates of the full supersymmetric Hamiltonian.
We can therefore use the fundamental property of
supersymmetric Hamiltonians that if $|\Psi\rangle$ is an eigenstate
with non-vanishing energy, then
$Q|\Psi\rangle$ is another eigenstate degenerate with the first.  Note
that
the state $Q|\Psi\rangle$ implicitly contains terms up to infinite
order in
$1/N$.  In our formalism, therefore, the concept of the supersymmetric
partner of a bosonic eigenstate only makes sense nonperturbatively.
Applying the full Hamiltonian to states $Q|\Psi\rangle$, contributions
of
order $1/N^2$ and higher fall away and we are left with the
energies \exactEnergies.

The above properties of the supersymmetric Hamiltonian follow formally
from the construction $H=\half\{Q,Q^\dagger\}$ along with the
properties $Q^2 = (Q^\dagger)^2 = 0$.  As a final check of the
correctness of the theory, one might verify in a perturbative context
that states $|\Psi\rangle$ and $Q|\Psi\rangle$ have the same energy up
to a given order in $1/N$.  To check that the above is true
up to order $1/N$ it is sufficient to show that $H_0 + {1\over N}H_3$
commutes with $Q_0 + {1\over N} Q_3$ up to corrections of order
$1/N^2$.  Explicitly, the oscillator expansion of $Q$ reads
$$
\eqalign{
  Q_0 &= 2\pi\sqrt{{\phi_0}\over L}\sum_{m\ne 0}\sqrt{|k_m|}
         \,b_m^\dagger a_m,  \cr
  Q_3 &= {1\over{L^3\phi_0}}\sum_{n, m \ne 0}\epsilon(n)
         \sqrt{|k_n k_m|}\left( b_{m+n}^\dagger a_n a_m
                -b_{m-n}^\dagger a_n^\dagger a_m^\dagger\right). \cr
 }  \eqno \eq
$$
Using the expressions \? and \oscillatorH, it is a straightforward
(though
tedious) exercise to verify that $H$ indeed commutes with $Q$ to
order $1/N$.

To summarise, by identifying the bosonic part of the supersymmetric
theory with the bosonic collective field theory, and using the fact
that the exact spectrum of the latter is known, we can in principle
use the supercharge to generate the exact spectrum of the
supersymmetric theory.  In practice this procedure is complicated by
the fact that the supercharge has an infinite perturbative expansion.

\section {Alternative representation of exact states}

\noindent
We now show that there exists a concise representation of
the fermionic partners of the exact bosonic states in terms of
the {\it unrescaled\/} fields of chapter 2.  This discussion is based
on work done in a different context by B. Sazdovi\'c [33].

The supercharge $Q$ in \continuumQ\ was given in terms of
the fields \DiscToCont\ in the free case as
$$
  Q = \int dx\,\psi^\dagger(x)\left(\sigma(x) -i\,\pv dy\,
         {\dely\eta\over x-y}\right),
  \eqno\eq
$$
where the fields satisfied the commutation relations
$$
 \eqalign{
  [\sigma(x), \eta(y)] &= -i\, \delta (x-y),  \cr
  [\sigma(x), \psi^\dagger(y)] &= i\, {{\psi^\dagger}\over\phi}(x)\,
                          \delx\delta(x-y),   \cr
  [\sigma(x), \psi(y)] &= i\, {{\psi}\over\phi}(x)\,
                          \delx\delta(x-y),   \cr
  \{\psi(x), \psi^\dagger(y)\} &= \phi(x)\,\delta(x-y),   \cr
 } \eqno \eq
$$
with $\phi \equiv \phi_0 + \partial\eta$.
We emphasize that we are now working with $\sigma$ and not $p$, and
that these $\psi$'s are not the ones used in the rest of the
chapter.  In terms of these fields the exact states
can be written in a particularly compact form  as follows [33].  On
$[-L,L]$ one can expand
$$
  \eqalign{
    \eta &\equiv \sqrt{1\over 4\pi L}\,\sum_n {1\over\sqrt{|k_n|}}
     \left( a_n e^{ik_n x}+ a^\dagger_n e^{-ik_n x}\right),   \cr
    \sigma &\equiv -i\sqrt{\pi\over 4L}\, \sum_n \sqrt{|k_n|}
        \left( a_n e^{i k_n x} - a^\dagger_n e^{-ik_n x}  \right),
        \cr
    \psi^\dagger &\equiv \sqrt{1\over 2L}\,\sum_n b^\dagger_n e^{-ik_n
       x},   \cr
    \psi &\equiv \sqrt{1\over 2L}\,\sum_n b_n e^{ik_n
       x},   \cr
  }  \eqno\eq
$$
where
$k_n\equiv \pi n/L$ and
 $[a_m, a_n] = \delta_{mn}$ as usual but $\{b_m, b^\dagger_n\}$
and $[b_m^{(\dagger)}, a_m^{(\dagger)}]$ are nontrivial.

In terms of these oscillators the supercharge has the simple form
$$
   Q = -i \sqrt{2\pi} \sum_{m\ne 0} \sqrt{|k_m|}\,b^\dagger_m a_m.
  \eqno\eq
$$
Also, from $[\psi^\dagger, \eta] = 0$ it follows that
$[b^\dagger_m, a^\dagger_n + a_{-n}] = 0$, which implies that
$$
\eqalign{
  [Q, a_n^\dagger + a_{-n}] &= -i \sqrt{2\pi|k_n|}\,b^\dagger_n \cr
          &= -i\sqrt{2\pi^2|n|\over L}\, b^\dagger_n.   \cr
}  \eqn\Qatob
$$

{}From [17] we know that one branch of exact states can be
written in terms of the so-called Schur polynomials $P_n$ as
$$
\eqalign{
  |\Psi_n\rangle &= P_n(\tilde a_1, \tilde a_2, \ldots)|0\rangle  \cr
     &\equiv {1\over 2\pi i} \oint dz\, z^{-n-1}
        \exp\left\{\sum_{n>0} \tilde a^\dagger_n z^n
          \right\}|0\rangle
        \cr
     &= {1\over 2\pi i} \oint dz\, z^{-n-1}
        \exp\left\{-i \sum_{n>0} {a^\dagger_n\over\sqrt n}\, z^n
             \right\}|0\rangle
              \cr
     &= {1\over 2\pi i} \oint dz\, z^{-n-1}
        \exp\left\{-i \sum_{n>0} {(a^\dagger_n+a_{-n})\over\sqrt n}
         \,   z^n   \right\}|0\rangle,
              \cr
}   \eqno\eq
$$
where $\tilde a^\dagger_n \equiv -i \epsilon(n)
a^\dagger_n/\sqrt{|n|}$.  The last step follows from the fact that the
$a_{-n}$ commute with the $a_n$ for $n>0$ and annihilate the vacuum.

Now using the fact that the combination $a^\dagger_n + a_{-n}$
tranforms
under $Q$ as in \Qatob\ and commutes with the $b_n$, one gets
$$
\eqalign{
  Q|\Psi_n\rangle &= {1\over 2\pi i}\oint dz\, z^{-n-1}
   \left(- \sum_{n>0}\sqrt{2\pi^2\over L}\,b_n^\dagger
z^n\right)\times
       \cr
   &\qquad \times
        \exp\left\{-i \sum_{n>0} {(a^\dagger_n+a_{-n})\over\sqrt n}
        \,z^n    \right\}|0\rangle.  \cr
}   \eqno\eq
$$
This can be written in a simpler form by notin that for
$\varepsilon$ Grassman one has
$$
\eqalign{
  e^{\varepsilon Q}|\Psi_n\rangle &= (1+\varepsilon Q)|\Psi_n\rangle
       \cr
   &= {1\over{2\pi i}}
        \exp\left\{\sum_{n>0}\left[
        (-i){(a^\dagger_n+a_{-n})\over\sqrt
              n} - \sqrt{2\pi^2\over L}\,\varepsilon b^\dagger_n
            \right] z^n  \right\}|0\rangle,  \cr
}  \eqno\eq
$$
so that
$$
  e^{\varepsilon Q}|\Psi_n\rangle
   = P_n (\tilde a^\dagger_1 + \tilde a_{-1} + \varepsilon \tilde
          b^\dagger_1,
          \tilde a^\dagger_2 + \tilde a_{-2} + \varepsilon \tilde
          b^\dagger_2, \ldots) |0\rangle,
  \eqno\eq
$$
where $\tilde b^\dagger_n\equiv - \sqrt {2\pi^2/L}\,b^\dagger_n$.

We can therefore write the superpartner of the state $|\Psi_n\rangle$
as
$$
\eqalign{
  Q|\Psi_n\rangle &= \int d\varepsilon\, e^{\varepsilon
          Q}|\Psi_n\rangle   \cr
     &= \int d\varepsilon\, P_n (\tilde a^\dagger_k + \tilde a_{-k}
                     + \varepsilon \tilde b^\dagger_k) |0\rangle   \cr
     &\equiv P_{S(n)}
                (\tilde a^\dagger_k + \tilde a_{-k},
                \tilde b^\dagger_k) |0\rangle,   \cr
}  \eqn\superSchur
$$
where $P_{S(n)}$ as defined above may be called a \lq\lq super-Schur
polynomial\rq\rq\ [33].

The above analysis can be extended to find the superpartner of a
general state, given by [17]
$$
  |\Psi\rangle_{(n_1,\ldots,n_N)} \equiv
      \xi_{(n_1,\ldots,n_N)}(\tilde a^\dagger_k + \tilde a_{-k})
    |0\rangle,     \eqno\eq
$$
where $\xi_{(n_1, \ldots, n_N)}$ is a general character polynomial
[30].  One finds
$$
  Q|\Psi\rangle_{(n_1,\ldots,n_N)} = \int d\varepsilon\,
      \xi_{(n_1,\ldots,n_N)}(\tilde a^\dagger_k + \tilde a_{-k}
        + \varepsilon \tilde b^\dagger_k)
    |0\rangle,     \eqno\eq
$$
in complete analogy with \superSchur.

In conclusion, we have obtained a compact expression for the exact
states in the supersymmetric theory.  This was achieved, however, in
a picture in which the oscillators have nontrivial commutation
relations with each other.

\section {Semiclassical analysis}

\noindent
We now address the possibility of reproducing nonperturbative effects
in a semiclassical analysis.
In the previous section we noted that the exact analysis of the
bosonic sector of the theory produces two \lq\lq single
particle\rq\rq\
branches \exactStates\ with dispersion relations \exactEnergies.  The
fact that
the corrections to the particle energies in \exactEnergies\ are of
order $1/N$
instead of the natural perturbation parameter $1/N^2$, implies that
these corrections are due to a nonperturbative effect.

The question of addressing this effect in a semiclassical analysis
for the bosonic string was
discussed in [17].  There an effective Lagrangian and Hamiltonian
were written down for each of the particle branches in \exactStates,
which
fit the exact dispersions given by equation \exactEnergies\ already at
the quadratic level.  The effective Hamiltonian for the first particle
branch was shown to be given by
$$
\eqalign{
  H_{{\rm eff}} &= \int dx\, \biggl\{ {1\over{2N^2}} \phi\,p^2
   +{1\over 8}{{\phi_{,x}^2}\over\phi}
    + {{N^2\pi^2}\over 6}\, \phi^3(x)   \cr
    &+ {N\over 2}\,\phi_{,x}\pv {dy\over{x-y}}\,\phi(y)\biggr\}. \cr
}  \eqn \effectiveH
$$

To extend this programme to the supersymmetric
case, we suggest that the proper effective Hamiltonian
to use would be the one constructed from the supercharges
$$
  \eqalign{
    Q &\equiv \int dx\,\psi^\dagger(x)\left({{\sigma(x)}\over N}
       - i N \pv  dy\,{{\dely\varphi}\over{x-y}}
       - {i \over 2} {{\phi_{,x}}\over\phi}
        \right),
    \cr
    Q^\dagger &\equiv \int dx\,\psi(x)\left({{\sigma(x)}\over N}
       + i N \pv dy\,{{\dely\varphi}\over{x-y}}
       + {i \over 2} {{\phi_{,x}}\over\phi}
        \right).  \cr
  } \eqn \effectiveQ
$$
The bosonic part of $H_{\rm eff}\equiv \half\{Q,Q^\dagger\}$ is
then easily seen to give the expression \effectiveH.

Whereas the original cubic bosonic collective field theory exhibited
nonperturbative states once exactly solved, the effective Hamiltonian
in equation \effectiveH\ exhibits these states at a
semiclassical level.  The first particle branch in \exactStates\
follows
immediately from the spectrum of the quadratic piece, while the second
branch was shown in [17] to correspond to nontrivial soliton
solutions of the effective Hamiltonian.

Within the context of the effective theory generated by the
supercharges in \effectiveQ, whose effective potential is given by
$$
\eqalign{
  V_{{\rm eff}} &= \int dx\, \biggl\{
   {1\over 8}{{\phi_{,x}^2}\over\phi}
    + {{N^2\pi^2}\over 6}\, \phi^3(x)
    + {N\over 2}\,\phi_{,x}\pv {dy\over{x-y}}\,\phi(y)\biggr\}, \cr
}  \eqno \eq
$$
one can identify instanton solutions by approximating the
dynamics of a
single real eigenvalue interacting with the field $\phi$ in the
potential \?.
Such single real eigenvalue configurations provide, within this
framework, a mechanism for supersymmetry breaking.

This concludes our discussion of the free theory, and we now move on
to a discussion of
the case of a harmonic superpotential.

\endpage

\chapter {THE HARMONIC CASE}

\section {Motivation}

\noindent
In this chapter, we study the model characterised by
the effective superpotential
$$
  W = \int dx\,\bar W(x)\,\delx\varphi - {1\over 2}
     \int dx \int dy\, \ln |x-y|\,\delx\varphi\,\dely\varphi.
     \eqno \eq
$$
in the case where $\bar W(x)={\omega\over 2}x^2$.  As remarked
before (see \generalContW), this model is a continuum version of the
original super-Calogero system \CalogeroW\ and \sCalogeroH.
Equivalently, it can be seen as a continuum formulation of a matrix
model of the form \matrixS, where the matrix superpotential
$\bar W(\Phi)={\omega\over 2}\Phi^2$ is purely
harmonic.  This system has been studied in the collective field
formalism before in [16], and an expanded version of their
analysis will be presented here.

\section {The vacuum density}

\noindent
We start the analysis by solving for the classical vacuum.  Due to the
fact that the potential is bounded, we shall see that, in contrast to
the free case, no regularization is needed at this level.  On the
other hand, due to turning point
singularities the fluctuations do have to be regularized.  This issue
will be discussed in sections (5.4) and (5.5).

Using $v(x) = W'(x) = \omega x$, the leading contribution to the
ground state energy of the
Hamitonian \perturbH\ is given by
$$
  V_{\rm eff}(\phi_0) = {{N^2} \over 2}\int dx\,\phi_0(x)
     \left(\pv dy\,{{\phi_0(y)} \over {x-y}} -
     \omega x\right)^2.
  \eqn \VeffHarmonic
$$
For supersymmetry to be
unbroken to leading order in $N$
the vacuum configuration of the density field must satisfy
$$
  V_{\rm eff}(\phi_0) = 0.
  \eqno \eq
$$
This will be the case if
$$
         \pv dy\,{{\phi_0(y)} \over
             {x-y}} - \omega x = 0,
  \eqn \vacCondition
$$
where $\phi_0$ must satify the constraint $\int\phi_0 = 1$.

The equation \vacCondition\ may be solved as follows : assume
$\phi_0$ has
support on some interval $(-a,a)$ and introduce the analytic function
$$
  F(z) \equiv \int_{-a}^a dy\,{\phi_0(y)\over x-y}  \eqno \eq
$$
defined for complex $z$ outside the interval $(-a,a)$.
By construction, $F$ is analytic in the complex plane with a cut along
$(-a,a)$, behaves as $1/z$ as $z\to\infty$ (due to the normalization
of $\phi_0$) and satisfies
$$
  F(x\pm i\varepsilon) = \omega x \mp i\pi\phi_0 (x)  \eqno\eq
$$
for $x\in (-a,a)$.
It is easy to check that
$$
  F(z) = \omega z - \omega\sqrt{z^2 - a^2}    \eqno\eq
$$
satisfies these conditions provided $a=\sqrt{2/\omega}$, which implies
that  $\phi_0$ is given in the interval $(-a,a)$ by
$$
  \phi_0 (x) = {1\over\pi}\sqrt {\mu_F - \omega^2 x^2};  \qquad
   \mu_F \equiv 2\omega.
  \eqn\harmPhizero
$$

We have found a solution to $V_{\rm eff} (\phi_0) = 0$
satisfying the constraint $\int\phi_0=1$. As
$V_{\rm eff}$ is positive definite, this solution is an extremum.
We can therefore conclude that for the harmonic superpotential the
vacuum energy is zero to leading order in the perturbation expansion.

Alternatively, one could have used the identity \unregId\ to
express the effective bosonic potential
$$
  V_{\rm eff}(\phi) = {{N^2} \over 2}\int dx\,\phi(x)
     \left(\pv dy\,{{\phi(y)} \over {x-y}} -
     \omega x\right)^2
  \eqno \eq
$$
of the Hamiltonian \rescaledHtwo in the form
$$
  V_{\rm eff}(\phi)
   = N^2 \left( {{\pi^2}\over 6}\int \phi^3
     -\omega\left(\int\phi\right)^2 - \omega^2 \int x^2\phi
         \right).   \eqno\eq
$$
The background $\phi_0$ satisfies the stationarity condition
$$
  {{\pi^2\phi_0^2}\over 2} - 2\omega \int\phi_0 + \omega^2x^2
   = 0.
  \eqn\harmStationarity
$$
Thus one gets
$$
  \phi_0 (x) = {1\over\pi}\sqrt {\mu_F - \omega^2 x^2},
  \eqno \eq
$$
where $\mu_F \equiv  2\omega\int\phi_0$.
Requiring that $\int\phi_0 = 1$, one finds $\mu_F = 2\omega$,
which is consistent with \harmPhizero.

This concludes our discussion of the vacuum configuration.  Expanding
about the classical vacuum, we can now move on to a determination of
the semiclassical spectrum.

\section{The quadratic spectrum}

\noindent
We now write the quadratic Hamiltonian in the momentum space
representation.  This will allow us to demonstrate supersymmetry
of the semiclassical spectrum.

Using the property
$$
         \pv dy\,{{\phi_0(y)} \over
             {x-y}} - \omega x = 0,
$$
and the fact that, due to the constraint, $\int dx\,\delx\eta = 0$,
one can write the quadratic piece of the Hamiltonian \perturbH\ in the
harmonic case $v(x) = \omega x$ as
$$
\eqalign{
  H_0 &=
       {1\over 2} \int dx\,\phi_0 p^2 + {{\pi^2}\over 2}\int dx\,
        \phi_0(\delx\eta)^2   \cr
    &+ {1\over 2}\int
          dx\,\left[\psi^\dagger(x)\sqrt{\phi_0(x)},
          {d\over{dx}}\pv
          dy\,{{\psi(y)\sqrt{\phi_0(y)}}\over{x-y}}\right].   \cr
}  \eqno\eq
$$

Defining the change of variables
$dq = dx/\phi_0$ and rescaling $p\to p/\phi_0$, $\psi \to
\psi/\sqrt{\phi_0}$, $\psi^\dagger\to\psi^\dagger/\sqrt{\phi_0}$, the
Hamiltonian can be rewritten as
$$
\eqalign{
  H_0 &=
       {1\over 2} \int dq\, p^2 + {{\pi^2}\over 2}\int dq\,
        (\delq\eta)^2   \cr
    &+ {1\over 2}\int
          dq\,\left[\psi^\dagger(q),
          {d\over{dq}}\pv
          dq'\,{{\phi_0(q')\psi(q')}\over{x(q)-x(q')}}\right].   \cr
}  \eqn\harmHzero
$$
The reason for choosing this reparametrization is that in $q$-space
the bosonic
part of the quadratic Hamiltonian reduces to that of a massless scalar
particle in a trivial background.  Although less obvious, the
fermionic part also has the spectrum of a massless particle,
as expected for a system with unbroken supersymmetry.
To confirm this property, we move on to a
calculation of the spectrum.

Explicitly, the above change of variables reads
$$
\eqalign{
  q &= \pi \int {dx\over\sqrt{\mu_F-\omega^2 x^2}}  \cr
    &= {\pi\over\omega}\arccos {-x\over a},
}   \eqno\eq
$$
where $q$ has been chosen to be zero at the turning point
$x = -a$.  The turning point was found in the previous section to be
given by $a = \sqrt{\mu_F}/\omega = \sqrt {2/\omega}$.
Inverting \?, one finds
$$
  x =
      -\sqrt{2L\over\pi^2} \cos{\pi q\over L}; \qquad q\in [0,L],
$$
where $L = {\pi^2/\omega} = \pi^2 a^2/2$ is the half period.

Expanding $\eta$ and $p$ on $[0,L]$ as
$$
\eqalign{
  \eta(q) &= \sum_{n=1}^\infty\, {1\over\sqrt{2\pi^2 n}}\,
          (a_n + a_n^\dagger)\, \sin{n\pi q\over L},  \cr
  p(q)  &= \sum_{n=1}^\infty\, -i\sqrt{\pi^2 n\over 2L^2}\,
          (a_n - a_n^\dagger)\, \sin{n\pi q\over L},  \cr
}   \eqno\eq
$$
where $[a_m,a_n^\dagger]=\delta_{mn}$,
the bosonic part of the quadratic Hamiltonian becomes
$$
  H_0{}^B = \sum_{n=0}^\infty\,n\omega\,(a_n^\dagger a_n + \half).
  \eqno\eq
$$

Explicitly, the fermionic part of the quadratic Hamiltonian reads
$$
 H_0{}^F
    = {\pi\over 2L}\int_0^L
          dq\,\left[\psi^\dagger(q),
          {d\over{dq}}\pv_{\!\!\!\! 0}^L
          dq'\,{{\sin {\pi
            q'\over L}\,\psi(q')}\over{\cos{\pi q'\over
                      L}-\cos{\pi q\over L}}}\right].
  \eqn\HzeroFermionic
$$
To find the spectrum one expands
$$
\eqalign{
   \psi(q) &= {1\over\sqrt L}\sum_{n=1}^\infty\, b_n\,
         \sin {n\pi q\over L},   \cr
   \psi^\dagger(q) &= {1\over\sqrt L}\sum_{n=1}^\infty\, b_n^\dagger\,
         \sin {n\pi q\over L},   \cr
}  \eqno \eq
$$
where $\{b_m, b^\dagger_n\}=\delta_{mn}$.

Inserting this expansion into \HzeroFermionic, the only nontrivial
part of the evaluation is the calculation of the integral
$$
\eqalign{
     \pv_{\!\!\!\! 0}^L
          dq'\,{{\sin {\pi
            q'\over L}\,\sin{\pi n q'\over L}}\over{\cos{\pi q'\over
                      L}-\cos{\pi q\over L}}}
     &= \half
     \pv_{\!\!\!\! -L}^L
          dq'\,{{\sin {\pi
            q'\over L}\,\sin{\pi n q'\over L}}\over{\cos{\pi q'\over
                      L}-\cos{\pi q\over L}}}   \cr
     &= \half\, {\rm Im}
     \pv_{\!\!\!\! -L}^L
          dq'\,{{\sin {\pi
            q'\over L}\,\exp{i\pi n q'\over L}}\over{\cos{\pi q'\over
                      L}-\cos{\pi q\over L}}},
}  \eqno \eq
$$
which can be done by closing the contour in the upper half plane,
exploiting the fact that the integrand is periodic with period $2L$.
The only relevant poles are on the real line at $+q$ and $-q$.  The
principal
value prescription corresponds to taking half of the residues of any
poles that lie on the contour of integration.  Applying this, one
obtains
$$
     \pv_{\!\!\!\! 0}^L
          dq'\,{{\sin {\pi
            q'\over L}\,\sin{\pi n q'\over L}}\over{\cos{\pi q'\over
                      L}-\cos{\pi q\over L}}}
   = - L \cos {\pi n q\over L}.
$$

Using this result, $H_0{}^F$ is now easily rewritten in the form
$$
  H_0{}^F = \sum_{n=1}^\infty \,n\omega\, (b^\dagger_n b_n-\half),
   \eqno \eq
$$
thus explicitly demonstrating supersymmetry of the semiclassical
spectrum.

\section {Three point functions}

\noindent
We now compute the supercharges and three point functions of the
quadratic theory in the oscillator basis.  This will be a naive
calculation, and questions of turning point regularization will be
postponed to the next section.

Using \vacCondition\ and changing variables to $q$-space as in the
previous section, the supercharges \perturbQ\ become
$$
 \eqalign{
   Q
     &=
        \int_0^L dq\, \psi^\dagger(q)
        \biggl\{
        p(q) - i\,\phi_0(q)\pv_{\!\!\!\! 0}^L
              dq'\,{{\delqp\eta}\over{x(q)-x(q')}} \biggr\} \cr
     &\quad + {1\over{2N}}\int_0^L
        {dq\over{\phi_0^2}}\,\psi^\dagger(q)\,
        \delq\eta\left\{
          p(q) - i\,\phi_0(q)
            \pv_{\!\!\!\! 0}^L
              dq'\,{{\delqp\eta}\over{x(q)-x(q')}}\right\} \cr
     &\quad - {i\over 2N}\int_0^L
        {dq\over{\phi_0^2}}\,\psi^\dagger
        (\delq\psi^\dagger)\psi
        + o\left({1\over N^2}\right),                 \cr
     &= \half
        \int_{-L}^L dq\, \psi^\dagger(q)
        \biggl\{
        p(q) - i\,\phi_0(q)\pv_{\!\!\!\! 0}^L
              dq'\,{{\delqp\eta}\over{x(q)-x(q')}} \biggr\} \cr
     &\quad + {1\over{4N}}\int_{-L}^L
        {dq\over{\phi_0^2}}\,\psi^\dagger(q)\,
        \delq\eta\left\{
          p(q) - i\,\phi_0(q)
            \pv_{\!\!\!\! 0}^L
              dq'\,{{\delqp\eta}\over{x(q)-x(q')}}\right\} \cr
     &\quad - {i\over 4N}\int_{-L}^L
        {dq\over{\phi_0^2}}\,\psi^\dagger
        (\delq\psi^\dagger)\psi
        + o\left({1\over N^2}\right),                 \cr
  Q^\dagger &= {\rm h.c.}, \cr
 }  \eqn \qspaceQ
$$
where we have used the fact that the fields satisfy Dirichlet boundary
conditions to extend the integral between $-L$ and $L$.

We now expand, as in the previous section
$$
\eqalign{
  \eta(q)
          &= {1\over\sqrt{2\pi L}}\sum_{n>0}\,
          {1\over \sqrt{|k_n|}}\,
          (a_n + a_n^\dagger) \sin k_n q,  \cr
          &= {1\over 2i}{1\over\sqrt{2\pi L}}\sum_{n\ne 0}\,
          {1\over \sqrt{|k_n|}}\,
          (a_n e^{ik_n q} + a_n^\dagger e^{-ik_n q}),  \cr
  p(q)  &= -{1\over 2}{\sqrt{\pi\over 2L}}\sum_{n\ne 0}\,
           \sqrt {|k_n|}\,
          (a_n e^{ik_n q} - a_n^\dagger e^{-ik_n q}),  \cr
   \psi(q) &= {1\over 2i}{1\over\sqrt L}\sum_{n\ne 0}\, b_n\,
         e^{ik_n q},   \cr
   \psi^\dagger(q) &= -{1\over 2i}{1\over\sqrt L}\sum_{n\ne 0}\,
         b^\dagger_n\,
         e^{-ik_n q},   \cr
}  \eqno \eq
$$
where $k_n = \pi n/L$ and where for convenience we have defined
$$
\eqalign{
a_{-n} &\equiv -a_n,  \qquad  a^\dagger_{-n} \equiv -a^\dagger_n,\cr
b_{-n} &\equiv -b_n,  \qquad  b^\dagger_{-n} \equiv -b^\dagger_n,\cr
}   \eqno \eq
$$
for $n>0$.

For the quadratic term in $Q$ one needs to compute terms of the type
$$
  \phi_0(q)\pv_{\!\!\!\! 0}^L
 dq'\,{{e^{ik_n q'}}\over{x(q)-x(q')}},
$$
where  $x=
 -\sqrt{2L\over\pi^2} \cos{\pi q\over L}$ and
$\phi_0 = x' = \sqrt{2\over L}\sin (\pi q/L)$.
This can easily be done via a contour integration similar to that in
the previous section.  One finds
$$
  \phi_0(q)\pv_{\!\!\!\! 0}^L
 dq'\,{{e^{ik_n q'}}\over{x(q)-x(q')}}
   = - \pi i\,\epsilon(k_n)\,e^{ik_n q}.  \eqn \cosId
$$

Using \cosId, the quadratic term becomes simply
$$
  Q_0 = -i\sqrt{\pi\over 2}\sum_{m>0}\, \sqrt{k_m}\, b^\dagger_m
  a_m.    \eqn \Qzero
$$

For the cubic terms in $Q$ one needs to compute the integrals
$$
  \pv_{\!\!\!\! -L}^L dq\, {e^{ik_n q}\over\phi_0^2(q)}.
$$
For $n>0$ this can be done by extending the contour in the upper half
plane vertically upwards at $-L$ and $L$. Exploiting the fact that the
integrand is periodic with period
$2L$, it follows that the vertical contributions cancel.  For $n<0$
one can similarly close the contour in the lower half plane. The
function
$1/\phi_0^2 = L / 2 \sin^2(\pi q/L)$ has singularities on the
contour at $-L$, at $0$ and at $L$.
Applying a principal part
regularization, which corresponds to taking half of the residues of
any poles
that lie fully on the contour and $1/4$ of the residues of poles that
are located at the corners $-L$ and $L$, one finds that for $n$ odd
the integral vanishes while for $n$ even it is given by
$$
  \pv_{\!\!\!\! -L}^L dq\, {e^{ik_n q}\over\phi_0^2(q)} =
    - {L^3\over\pi}\, |k_n|.     \eqn\kernelB
$$

Using this result one finds, after some algebra, that the
cubic piece of $Q$ is given by
$$
\eqalign{
  Q_3 &= {iL\sqrt L\over 8\pi N}\,
    {\sum_{m,n,p\ne 0}}{}^{\!\!\!\!\! \prime}\,
\epsilon(n)\sqrt{|k_nk_p|}\,|k_m+k_n+k_p|
      \,(b^\dagger_{-m}a_n a_p- b^\dagger_{-m}a^\dagger_{-n}a_p) \cr
  &+ {iL\sqrt L\over 8\pi N}\,
    {\sum_{m,n,p\ne 0}}{}^{\!\!\!\!\! \prime}\,k_n|k_m+k_n-k_p|\,
      b^\dagger_m b^\dagger_n b_p,    \cr
}  \eqn\Qthree
$$
where the primes on the sums indicate that one only sums over
indices such that the arguments of the absolute value signs are even.

The cubic piece of the Hamiltonian can now be generated from \Qzero\
and \Qthree\ using $H = \half\{Q,Q^\dagger\}$.  One finds, after some
algebra, that
$$
  H_3 = H_3{}^B + H_3{}^F,
$$
where
$$
\eqalign{
  H_3{}^B &= {L\over 4N}\sqrt{L\over 2\pi}\,
            {\sum_{n,p,l>0}}{}^{\!\!\!\!\! \prime}\,\sqrt{k_l k_n k_p}\,
            \bigl(|k_l+k_n+k_p|    \cr
          &-|k_l-k_n-k_p|\bigr)\,
            a^\dagger_n a^\dagger_p a_l + {\rm h.c.},   \cr
  H_3{}^F &= {L\over 8N}\sqrt{L\over 2\pi}\,
             {\sum_{n,p,l>0}}{}^{\!\!\!\!\! \prime}\,\sqrt{k_l}\,(k_p-k_n)\,
             \bigl(|k_l+k_n+k_p|-|k_l+k_n-k_p|  \cr
          &\quad\,  +|k_l-k_n+k_p|-|k_l-k_n-k_p|\bigr)\,
             b_n^\dagger b_p a_l^\dagger + {\rm h.c.}   \cr
}   \eqn \Hthree
$$

These expressions have been calculated using a principal part
prescription to regularize turning point divergences.  In the next
chapter we shall present a more careful discussion of the
regularization.

\section {Turning point divergences}

\noindent
We now discuss the regularization of turning point divergences as
it applies to the calculation of the three point functions of the
previous section.

In chapter 3 it was mentioned that there may in general be subtleties
in the application of the identity
$$
 \int dx\,\delx\eta
   \left(\pv dy\,{\dely\eta \over {x-y}}\right)^2
 = {{\pi^2}\over 3}\int dx\,(\delx\eta)^3
\eqn\xspaceId
$$
given in $q$-space by
$$
   \int {dq}\,\delq\eta
            \left(\pv_{\!\!\!\! 0}^L
              dq'\,{{\delqp\eta}\over{x(q)-x(q')}}\right)^2   =
   {\pi^2\over 3}\int {dq\over\phi_0^2}\,(\delq\eta)^3,
\eqn\qspaceId
$$
used to simplify the
cubic piece of the Hamiltonian.

Indeed, in chapter 4 we saw that in the potential free case we had to
introduce a periodic regularization in order to make the right hand
side of \xspaceId\ convergent.

In the present case, due to the fact that the potential is bounded,
there are no problems in applying \xspaceId\ or \qspaceId\ to the the
classical vacuum $\phi_0$.  However, at the level of the fluctuations
$\partial\eta$, turning point divergences will arise on the right
hand side of the identity and have to be suitably regulated.  This was
already realized in [6] in the context of the bosonic collective
field theory.

In the calculation of the oscillator expansions
in the previous section, we effectively started from
the left hand side of \qspaceId.  Our regularization scheme
consisted of naively taking principal parts at the turning points.
On the other hand, the authors of [6] started from
the right hand side of \qspaceId, their regularization also coming
down to taking principal
parts. Comparing the bosonic part of our Hamiltonian \Hthree\
to the one that
was obtained in [6] for the bosonic
collective field theory, we see that our expression differs from
theirs in that we have no terms of the form $a_m a_n a_p$.

To see this difference more clearly, note that
starting from the left hand side of \qspaceId,
assuming that $\eta$ obeys Dirichlet boundary conditions so that we
can expand $\delq\eta = \sum_{n>0}\,\phi_n\cos k_n q$, and using
the naive principal part prescription, one finds
$$
\eqalign{
   \int {dq}\,\delq\eta
            &\left(\pv_{\!\!\!\! 0}^L
              dq'\,{{\delqp\eta}\over{x(q)-x(q')}}\right)^2  \cr
   &= -{\pi L^3\over 24} \sum_{mnp>0}{}^{\!\!\!\!\!
       \prime}\,\phi_m\phi_n\phi_p\,
         \bigl(-3|k_m+k_n+k_p|+|k_m-k_p-k_n|  \cr
   &\quad\, +|k_m+k_p-k_n|
             +|k_m-k_p+k_n|\bigr),
}   \eqn\naiveExpansion
$$
where the result \kernelB\ has been used.  Starting from the right
hand side of \qspaceId\ as in the bosonic collective field theory,
one finds
$$
\eqalign{
   {\pi^2\over 3}\int &{dq\over\phi_0^2}\,(\delq\eta)^3   \cr
   &= -{\pi L^3\over 24} \sum_{mnp>0}{}^{\!\!\!\!\!
        \prime}\,\phi_m\phi_n\phi_p\,
         \bigl(|k_m+k_n+k_p|+|k_m-k_p-k_n|    \cr
   &\quad\, +       |k_m+k_p-k_n|
          +|k_m-k_p+k_n|\bigr),           \cr
}   \eqn\cftExpansion
$$
which is manifestly different from the left hand side.

It is therefore obvious that
our naive regularization scheme differs from the one used in
the bosonic collective field theory.  After
indicating the origin of this difference, we will argue that
the regularization used in the bosonic collective field theory seems
to be inconsistent with supersymmetry.

The reason for the difference between \naiveExpansion\ and
\cftExpansion\ is as follows:  In [6] the turning point divergence
in \cftExpansion\ was regulated essentially by introducing a cutoff
$\epsilon$ so that the integration range becomes
$\int_\epsilon^{L-\epsilon}$.  Keeping epsilon small but finite
throughout and
only at the end discarding all $\epsilon$-dependent quantities (which
are argued to be nonuniversal), one obtains the principal part
prescription.   Applying the same cutoff to \naiveExpansion\
corresponds to identifying the $\epsilon$-independent terms in
$$
\int_\epsilon^{L-\epsilon}dq\,\dots\left(\pv_{\!\!\!\!
\epsilon}^{L-\epsilon}
      dq'\,\dots
   \right)^2.  \eqno\eq
$$
As it stands, in \naiveExpansion\ we calculated not the quantities \?
but instead
$$
\int_\epsilon^{L-\epsilon}dq\,\dots\left(\pv_{\!\!\!\! 0}^{L}
      dq'\,\dots
   \right)^2.  \eqno\eq
$$

If we want correspondence with the bosonic cubic collective field
theory, we must therefore subtract
corrections of the typical form
$$
\int_\epsilon^{L-\epsilon}dq\,\dots\left(\pv_{\!\!\!\! 0}^{L}
      dq'\,\dots
   \right)
    \left(\int_{0}^{\epsilon}
      dq''\,\dots
   \right)  \eqno\eq
$$
from \naiveExpansion.

Explicitly, using \cosId\ and remembering that $x \propto \cos \pi
q/L$ and $\phi_0 \propto \sin \pi q/L$, we have
$$
\eqalign{
   \int_\epsilon^{L-\epsilon} & dq\,\cos k_m q \,\pv_{\!\!\!\! 0}^{L}
      dq'\,{\cos k_n q'\over x(q) - x(q')}
   \, \int_{0}^{\epsilon}
      dq''\,{\cos k_p q''\over x(q)-x(q'')}
       \cr
  &\propto
   \int_\epsilon^{L-\epsilon} dq\,{\cos k_m q \sin k_n q \over
      \phi_0(q)}
    \,\int_{0}^{\epsilon}
      dq''\,{\cos k_p q''\over x(q)-x(q'')}
       \cr
  &\sim
   \int_\epsilon^{L-\epsilon}  dq\,{\cos k_m q \sin k_n q \over
      \sin {\pi q\over L}}
    \epsilon\left(
      {1\over \cos {\pi q\over L}-1}
   \right).    \cr
}  \eqn\approx
$$
To see how we obtain a finite contribution from this, note that one
gets
an $o(\epsilon)$ contribution from the integral $\int_0^\epsilon dq''$
and an $o(1/\epsilon)$ contribution (see below) from each of the
limits of the integral $\int_\epsilon^{L-\epsilon} dq$.  The
$\epsilon$-dependence will therefore cancel.  More precisely, the
small $q$ behaviour of the integrand in \approx\ is given by
$$
  {\cos k_m q}\,{\sin {n\pi q\over L}\over\sin{\pi q\over L}}\,
      {1\over \cos {\pi q\over L}-1}
  \sim {n\over \half {\pi^2 q^2\over L^2}}
  \propto {k_n\over q^2}.
$$
Inserting this into the integral in \approx, one gets contributions of
the form
$$
  \epsilon \int_\epsilon dq\,{k_n\over q^2}
  \sim \epsilon\, k_n\, {1\over\epsilon} \sim k_n.
$$

We therefore see that these subtractions give rise to finite (\ie,
$\epsilon$-independent) corrections to the expansion \naiveExpansion\
proportional to
$$
  \sum_{mnp>0} k_n\,\phi_m\phi_n\phi_p =
  {1\over 3}\sum_{mnp>0} (k_m +k_n+k_p)\,\phi_m\phi_n\phi_p.
   \eqno\eq
$$
These corrections are exactly of the right form and the right sign to
restore equality between \naiveExpansion\ and \cftExpansion.  There
are four such terms, corresponding to the upper and the lower limit of
the integrations over each of $q'$ and $q''$.  Due to the crudeness
of the approximation in \approx, the above argument is not expected
to give the correct overall coefficient of the correction term.
However, this coefficient can be inferred from the fact that for a
finite cutoff $\epsilon$, all integrals are well behaved and the
identity \qspaceId\ will be valid.  We therefore expect the correction
term to be exactly equal to the difference between \naiveExpansion\
and \cftExpansion, which is given by
$$
   -{\pi L^3\over 24} \sum_{mnp>0}{}^{\!\!\!\!\!
       \prime}\,4\,(k_m+k_n+k_p)\,\phi_m\phi_n\phi_p.
$$

Once the coefficients have been determined, we can can work backwards
and write down the {\it regulated\/} expressions
$$
\eqalign{
  \int dq\, {\cos k_m q \sin k_n q \over \phi_0^2}
     &\left(\phi_0 \pv dq'' {\cos k_p q''\over x(q) - x(q'')}\right)
   \cr
  &= - {L^3\over 8}\bigl(|k_m+k_n-k_p| + |k_m-k_n+k_p|
     \cr
  &\quad -|k_m+k_n+k_p| - |k_m-k_n-k_p| + 2k_n    \bigr)
}   \eqno\eq
$$
and
$$
\eqalign{
  \int dq\, {\cos k_m q \over \phi_0^2}
     &\left(\phi_0 \pv dq' {\cos k_n q'\over x(q) - x(q')}\right)
     \left(\phi_0 \pv dq'' {\cos k_p q''\over x(q) - x(q'')}\right)
   \cr
  &= - {\pi L^3\over 8}\bigl(|k_m+k_n-k_p| + |k_m-k_n+k_p|
     \cr
  &\quad -|k_m+k_n+k_p| - |k_m-k_n-k_p| + 2k_n + 2k_p   \bigr),
}   \eqno\eq
$$
where the final terms are cutoff independent corrections to
the naive expressions obtained from the principal part prescription.

Using the second of these two equalities in the calculation of the
expansion of
\naiveExpansion, we indeed find equivalence between \naiveExpansion\
and \cftExpansion.

The first equality can be used in the calculation
of the cubic part of the supercharge $Q$ of the previous section.
However, and we want to stress this point, this regulation is
{\it inconsistent\/} with supersymmetry in the following sense:
discarding cutoff-dependent terms first before and then  after
calculating
$H=\half\{Q,Q^\dagger\}$, one finds inequivalent expressions for $H$.
This can be traced to cutoff dependent contributions
to the quadratic part of $Q$ combining with cutoff
dependent
contributions to the cubic part of $Q^\dagger$ to give finite cubic
terms in $H$.
In other words, the regularization does not commute with taking the
bracket.

The interpretation of this result is not clear.  It may be an
indication
that it is impossible to define a supersymmetric theory of which the
bosonic part agrees with the bosonic collective field theory of
[6].  On the other hand, the message may well be that some
kind of dynamical supersymmetry breaking is at work.
This is a topic that needs further investigation.

If we are willing to overlook the lack of correspondence with
the bosonic collective field theory,
we have the option of simply regulating the integrals according
to the naive prescriptions of the previous section.  The naive
regularization commutes with the operation $H=\half\{Q,Q^\dagger\}$,
and can therefore be taken to define a supersymmetric theory.

This concludes our discussion of the harmonic theory.  In the next
chapter we move on to more general potentials.

\endpage

\chapter {GENERAL POTENTIAL}

\noindent
In the previous two chapters we discussed the potential free case
$v(x) = 0$ and the harmonic case $v(x) = \omega x$.  We now generalise
the discussion to the case where $v$ can be a more general polynomial
in $x$.  We will start by demonstrating supersymmetry of the quadratic
spectrum at the semiclassical level.  We then move on to a discussion
of the criticality of the potential, which will be seen to correspond
to that of $c<1$ zero-dimensional matrix models.  We conclude with a
discussion of the prospects for finding a suitable scaling
limit for defining a $d=1$ superstring theory.

\section {Semiclassical spectrum}

\noindent
In the previous two chapters supersymmetry of the quadratic spectrum
was explicitly demonstrated in the potential free case and in the
harmonic case.  In this section we generalise this result to the case
of an arbitrary potential $v(x)$. In particular, we show that if
there exists a supersymmetric classical vacuum density, the
quadratic spectrum will be supersymmetric.

Referring back to chapter 3, we note that
the condition \nonFreeCond\ for supersymmetry at the level of the
classical vacuum  can equivalently be restated as
$$
  \left. W_{;x} \right|_{\phi_0} = 0,  \eqno\eq
$$
where $W$ stands for the effective superpotential of
\generalContW.
When this condition is satisfied,
the continuum Hamiltonian \continuumH\ can
be expanded around $\phi_0$ to give a quadratic contribution
$$
 H_0 = H_0^B + H_0^F,
\eqno\eq
$$
where
$$
\eqalign{
  H_0{}^B &= \half \int dx\,\phi_0 (x)\left(p^2
   + \int dy\int dz\, W_{;xy}W_{;xz}\,\eta(y)\,\eta(z)\right), \cr
  H_0{}^F &= \half\int dx\int dy\,
    \left[\psi^\dagger(x), \sqrt{\phi_0(x)}\, W_{;xy}\,
          \sqrt{\phi_0(y)} \psi(y)
  \right],    \cr
}   \eqno \eq
$$
and where the expressions of the form $W_{;xy}$ are evaluated at
$\phi_0$.

Defining the change of variables
$dq = dx/\phi_0$ and rescaling $p\to p/\phi_0$, $\psi \to
\psi/\sqrt{\phi_0}$, $\psi^\dagger\to\psi^\dagger/\sqrt{\phi_0}$, the
Hamiltonian simplifies to give
$$
\eqalign{
  H_0{}^B &= \half \int dq\,\left(p^2
   + \left(\int dq'\; W_{;qq'}\,\eta(q')\right)^2\right), \cr
  H_0{}^F &= \half\int dq\,
    \left[\psi^\dagger(q), \int dq'\, W_{;qq'}\,
           \psi(q')
  \right],            \cr
}   \eqno \eq
$$
where we have used the identity $\delta/\delta\varphi(q) =
\phi_0(q)\,\delta/\delta\varphi(x)$, which follows by the chain rule,
keeping in mind the fact that $\delta (q-q') = \phi_0
(q)\,\delta(x-x')$.

Now let $\tilde W$ be the kernel defined by
$$
\tilde W (\phi)(q) \equiv \int dq'\,W_{;qq'\,}\phi(q').  \eqno\eq
$$
Then the quadratic Hamiltonian can be written as
$$
\eqalign{
  H_0{}^B &= \half \int dq\,\left(p^2
   + \left(\tilde W(\eta)(q)\right)^2\right), \cr
  H_0{}^F &= \half\int dq\,
    \left[\psi^\dagger(q), \tilde W
           (\psi)(q)
  \right],            \cr
}   \eqn \kernelH
$$

Let $\{\phi_n\}$ be a complete set of normalised eigenfunctions of
$\tilde W$, \ie,
$$
  \tilde W \phi_n = \lambda_n\phi_n.  \eqno \eq
$$
Assuming $\tilde W$ to be positive definite, we can expand the fields
as
$$
\eqalign{
  \eta &= \sum_n\, {1\over\sqrt{2\lambda_n}}\,
           (a_n \phi_n + a^\dagger_n \phi^*_n),   \cr
  p    &= \sum_n\, -i\sqrt{\lambda_n\over 2}\,
           (a_n \phi_n - a^\dagger_n \phi^*_n),   \cr
  \psi &= \sum_n\, b_n\phi_n,   \cr
  \psi^\dagger &= \sum_n\, b^\dagger_n\phi^*_n,   \cr
}  \eqno \eq
$$
where $[a_m, a^\dagger_n] = \delta_{mn}$ and $\{b_m,
b^\dagger_n\}=\delta_{mn}$, all other commutators vanishing.

It is then trivial to show that
$$
\eqalign{
  H_0{}^B &= \sum_n\, \lambda_n (a_n^\dagger a_n + \half),  \cr
  H_0{}^F &= \sum_n\, \lambda_n (b_n^\dagger b_n - \half),  \cr
}  \eqno \eq
$$
thus demonstrating explicit supersymmetry of the quadratic spectrum.

It should be emphasized that the above argument is only valid in the
case
where there exists a supersymmetric classical configuration
$\phi_0$, and does
not make any
statement on the presence or absence of supersymmetry at the quadratic
level when one expands about an extremum that is not
supersymmetric.

Now, as we already noted in chapter 3, for a certain subclass of
potentials $v(x)$, certain simplifications occur.  To investigate
this, note that
the bosonic part of the quadratic piece in the perturbation expansion
\perturbH\ of the Hamiltonian was found to be given by
$$
  H_0{}^B = {1\over 2} \int dx\,\phi_0 p^2 + {{\pi^2}\over 2}\int dx\,
        \phi_0(\delx\eta)^2
         -\half\int dx\pv dy\,{v(x)-v(y)\over
           x-y}\,\delx\eta\,\dely\eta.
\eqno\eq
$$
When $v$ is of the general form
$v(x) = a + bx + cx^2$, the third term in this expansion falls away
due to the constraint $\int dx\, \delx\eta = 0$, and the quadratic
spectrum reduces to that of a massless scalar.  Explicitly, one has in
$q$-space
$$
  H_0{}^B = {1\over 2} \int dq\,\left( p^2 + {{\pi^2}}
        (\delq\eta)^2\right).   \eqno\eq
$$

In the light of the above discussion, we already know that assuming
that there exists a supersymmetric classical configuration, the
fermionic spectrum will also be that of a massless scalar particle.
One should therefore be able to rewrite the fermionic piece
$$
  H_0{}^F = {1\over 2}\int
          dx\,\left[\psi^\dagger(x)\sqrt{\phi_0(x)},
          {d\over{dx}}\pv
          dy\,{{\psi(y)\sqrt{\phi_0(y)}}\over{x-y}}\right]
\eqno\eq
$$
in a form in which this property is manifest.

Assuming $q$ to be defined on $[0, L]$, we expand
$\eta$ and $p$ as
$$
\eqalign{
  \eta(q) &= \sum_{n=1}^\infty\, {1\over\sqrt{2\pi^2 n}}\,
          (a_n + a_n^\dagger)\, \sin{n\pi q\over L},  \cr
  p(q)  &= \sum_{n=1}^\infty\, -i\sqrt{\pi^2 n\over 2L^2}\,
          (a_n - a_n^\dagger)\, \sin{n\pi q\over L},  \cr
}   \eqn\sineExpansion
$$
where $[a_m,a_n^\dagger]=\delta_{mn}$.
In the above Dirichlet boundary conditions have been assumed.

The bosonic part of the quadratic Hamiltonian is then simply given by
$$
  H_0{}^B = \sum_{n=0}^\infty\,{n\pi^2\over L}\,(a_n^\dagger a_n +
\half).
  \eqn\sineH
$$

In \kernelH\ the Hamiltonian was expressed in terms of a kernel
$\tilde W$ as
$$
\eqalign{
  H_0{}^B &= \half \int dq\,\left(p^2
   + \left(\tilde W(\eta)(q)\right)^2\right), \cr
  H_0{}^F &= \half\int dq\,
    \left[\psi^\dagger(q), \tilde W
           (\psi)(q)
  \right].            \cr
}   \eqno \eq
$$
The same kernel therefore appears in both the bosonic and the
fermionic part of the Hamiltonian.
Remembering that we want to rewrite the fermionic piece in a simpler
form, our motivation is therefore to solve for $\tilde W$ by
determining its properties.

Expanding with respect to a complete set of normalised
Dirichlet eigenfunctions of $\tilde W$ on $[0,L]$, we had
$$
\eqalign{
  \eta &= \sum_n\, {1\over\sqrt{2\lambda_n}}\,
           (\tilde a_n \phi_n + \tilde a^\dagger_n \phi^*_n),   \cr
  p    &= \sum_n\, -i\sqrt{\lambda_n\over 2}\,
           (\tilde a_n \phi_n - \tilde a^\dagger_n \phi^*_n),   \cr
}  \eqn \tildeExpansion
$$
and
$$
\eqalign{
  H_0{}^B &= \sum_n\, \lambda_n (\tilde a_n^\dagger \tilde a_n +
\half). \cr
}  \eqn\tildeH
$$

Comparing \tildeH\ and \sineH, one immediately sees that in the
present case $\tilde W$ has eigenvalues $\lambda_n = \pi^2 n/L$.
All that remains is therefore to explicitly solve for its
eigenfunctions $\phi_n$.

As both \sineExpansion\ and \tildeExpansion\ are expansions of
the same fields in terms of normalised oscillator coordinates, it
follows that the $a$'s are unitarily related to the $\tilde a$'s,
\ie,
$$
  a_n = U_{ni} \tilde a_i,  \qquad a^\dagger_n = U^*_{ni}
    \tilde a^\dagger_i,   \eqno\eq
$$
where $U^\dagger U=1$.  Comparing \tildeH\ and
\sineH\ one finds
$$
 \sum_n\, \lambda_n \tilde a_n^\dagger \tilde a_n
 = \sum_{n\,i\,j}\, \lambda_n\, U^*_{nj}\,\tilde a_j^\dagger\,
     U_{ni}\,\tilde a_i,
  \eqno\eq
$$
from which it follows by comparing coefficients that
$U^\dagger\Lambda U = \Lambda$, where $\Lambda$ is the diagonal matrix
with $\Lambda_{ii} = \lambda_i$.  Using the fact that $U$ is unitary,
this is equivalent to $[U, \Lambda] = [U^\dagger, \Lambda]= 0$.  Thus
$U$ leaves the eigenspaces of $\Lambda$ invariant, and as the
eigenvalues $\lambda_n = \pi^2 n/L$ are nondegenerate, we conclude
that $U$ is diagonal with complex phase factors on the diagonal.

Thus each $a_n$ is related to $\tilde a_n$ by a
phase factor.  Comparing \sineExpansion\ and \tildeExpansion, it
follows that the eigenfunctions of $\tilde W$ are given by
$$
  \phi_n(q) = {1\over\sqrt L} \sin {\pi n q\over L},
\eqno\eq
$$
modulo an inessential phase. In other words,
$$
  \tilde W \sin {\pi n q\over L} = {\pi^2 n\over L}\sin {\pi n q\over
   L}. \eqno\eq
$$

Using the identity
$$
 \pv dq\, {e^{ikq}\over q'-q}= -\pi i\,\epsilon(k)\,e^{ikq'},
\eqno\eq
$$
it is a simple matter to verify that
$$
 {d\over dq'}\pv dq\, {\sin{kq}\over q'-q}= \pi \,|k|\,
       \sin{kq'}, \eqno\eq
$$
so that the kernel can be written as
$$
  \tilde W (\phi) (q') =
 {d\over dq'}\pv dq\, {\phi{(q)}\over q'-q}
        \eqno\eq
$$

The conclusion is therefore that the fermionic piece
of the Hamiltonian \kernelH\ can be written as
$$
\eqalign{
  H_0{}^F &= \half\int dq\,
    \left[\psi^\dagger(q), \tilde W
           (\psi)(q)
  \right],            \cr
   &= \half \int dq\,
    \left[\psi^\dagger(q),
           {d\over dq}\pv dq'\, {\psi{(q')}\over q-q'}
  \right],            \cr
}   \eqno \eq
$$
or, in terms of the \lq\lq absolute\rq\rq\ derivative, defined in
section (4.4),
$$
\eqalign{
  H_0{}^F &= {-\pi i\over 2}\int dq\,
    \left[\psi^\dagger, |\delq|
           \psi
  \right].            \cr
}   \eqno \eq
$$
This concludes the demonstration that with the above restriction on
the potential $v(x)$, the quadratic
Hamiltonian can be rewritten in a form manifestly that of a free
Majorana fermion.  On the way we have proved the identity
$$
\int
   dq\,\psi^\dagger(q)\,
   {d\over{dq}}\pv
   dq'\,{{\phi_0(q')\psi(q')}\over{x(q)-x(q')}}
= \int dq\,
  \psi^\dagger(q) \,
       {d\over dq}\pv dq'\, {\psi{(q')}\over q-q'},
\eqno\eq
$$
valid when the potential $v(x)$ is at most quadratic in $x$.  This
identity may look deceptively trivial to prove via contour integration
arguments, due to the fact that $\phi_0 = x'$, so that both integrands
have the same residues.  However, in general the function $x$ may have
additional poles in unfavourable positions for the naive argument to
be valid.

As already mentioned in chapter 3, for $v(x)$ of cubic and higher
order in $x$, one obtains an additional nonlocal quadratic
contribution given by
$$
         -\half\int dx\pv dy\,{v(x)-v(y)\over
           x-y}\,\delx\eta\,\dely\eta.
$$
This term introduces an effective mass for the scalar boson and, by
supersymmetry, also for the Majorana fermion.

\section {Scaling limit}

\noindent
Ultimately we are interested in the large $N$ behaviour of the system
under consideration, a fact that was already implicit when we
approximated discrete densities by continuous functions.  In
particular, we would like to know whether a suitable scaling limit
exists and
what physical interpretation to attach to it.  With this in mind, we
now turn to a discussion of the scaling behaviour of the potential.

Consider the bosonic part of the potential in \rescaledHtwo:
$$
  V_{\rm eff}(\phi) = {{N^2} \over 2}\int dx\,\phi(x)
     \left(\pv dy\,{{\phi(y)} \over {x-y}} - v(x)\right)^2,
  \eqno \eq
$$
where $v(x)$ is an arbitrary polynomial.
A sufficient condition for supersymmetry to be preserved to leading
order in $N$ is that there exist a classical configuration $\phi_0$
such that
$$
         \pv dy\,{{\phi_0(y)} \over
             {x-y}} - v(x) = 0,
  \eqn \condA
$$
with $\int\phi_0 = 1$.
There is a well-known way of solving this equation using analytic
methods [31].

Alternatively, assuming that  $v$ has been chosen in such
a way as to avoid
possible problems with the identity \unregId\
such as occur in the free case (section (4.1)),
we can rewrite the bosonic potential in the form
$$
 \tilde  V_{\rm eff}(\phi)/N^2 = {\pi^2\over 6}\int \phi^3
      + \half\int v^2 \phi - \mu \left(\int \phi - 1\right)
      - \half\int dx \pv dy \,{v(x)-v(y)\over x-y}\,\phi(x)\,\phi(y),
  \eqno \eq
$$
where the Lagrange multiplier $\mu$ has been introduced to enforce the
constraint $\int\phi = 1$.
Classical vacua can be found by extremizing this potential with
respect to $\phi$.
The condition
for $\phi_0$ to be an extremum of \? is given by
$$
  {\pi^2\over 2}\, \phi_0^2(x)
      + \half\, v^2(x)  - \mu
      - \pv dy \,{v(x)-v(y)\over x-y}\,\phi_0(y) = 0,
  \eqn \condB
$$
where $\mu$ is determined in terms of the other parameters of the
theory by the above constraint.

Assuming that the vacuum configuration is supersymmetric, \ie,
$V_{\rm eff} (\phi_0) = 0$, the Lagrange multiplier will be zero.
This is a simple consequence of the fact that in such a case $\phi_0$
is an genuine minimum of the (positive definite) potential $V_{\rm
eff}$ rather than just a minimum of
$V_{\rm eff}$ relative to the constraint.  To see this more clearly,
note that
the condition of stationarity with respect to variations in $\phi$,
given by
$$
\delta_\phi \tilde V_{\rm eff}
    = \delta_\phi V_{\rm eff} - \mu = 0,   \eqno \eq
$$
implies that $\mu=0$ when $\delta V_{\rm eff} = 0$, which will indeed
be the case at $\phi_0$ satisfying $V_{\rm eff}(\phi_0) = 0$.
Conversely, if $\mu\ne 0$ then $V_{\rm eff}(\phi_0) \ne 0$, and
supersymmetry is broken.   The stationarity
condition \condB\ is therefore more general than the condition \condA\
in that \condB\ may allow us to find extrema that
are not supersymmetric, \ie, such that $V_{\rm eff}(\phi_0) \ne 0$,
in which case we may in general have
$\mu \ne 0$.  In other words, while
\condA\ is sufficient for stationarity, the second condition \condB\
is necessary and sufficient.

When the classical vacuum configuration is supersymmetric then, even
though the Lagrange multiplier
$\mu=0$, the last term in \condB\ may generate an effective chemical
potential, an example of which we saw in the harmonic case (see
equation \harmStationarity\ and the subsequent remarks).

It is at present unclear what the spacetime interpretation of the
one-dimensional supersymmetric matrix model should be.  In the $d=1$
matrix model formulation of the $c=1$ bosonic string,
an extra dimension is dynamically generated
[2] to give an effective two-dimensional theory, the extra
dimension being associated to the Liouville degree of freedom.  The
essential ingredient in the derivation is the existence of a double
scaling limit in which the turning point of the vacuum configuration
appoaches a maximum of the potential, in the vicinity of which the
potential behaves as $-x^2$.  As the maximum is approached, the
\lq\lq time of flight\rq\rq\ becomes infinite, hence the extra
dimension.

Let us now consider the scaling behaviour of the supersymmetric
theory.
We start by
noting that equation \condA\ or \condB\ (for $\mu = 0$) defining a
supersymmetric vacuum configuration is identical to that found in
zero-dimensional matrix models  with potential $w(x)$ such that
$v = w'$ [31], as also noted in [32].
This should not be surprising, as the original motivation for the
introduction of
the Marinari-Parisi supersymmetric matrix model was to stabilize the
critical behaviour of $d=0$ matrix models having unbounded potentials.

In the double scaling limit these $d=0$ models are known to describe
two-dimensional quantum gravity coupled to
nonunitary minimal matter of central charge $c<1$, the value of $c$
depending on the order of multicriticality, \ie, the details of the
behaviour
of the effective potential as one approaches the critical point.
This is in
contrast to the bosonic case, where the generation of the extra
dimension is related to the fact that the scaling of the potential
corresponded to $c=1$,  and we therefore expect the
spacetime interpretation of the supersymmetric matrix model not to be
as neat as in the bosonic case.

To illustate the difference between the bosonic and the fermionic
scalings, consider first the bosonic analogue of the stationarity
condition \condB.  It is given by
$$
  {\pi^2\over 2}\, \phi_0^2(x)
      +  w(x)  - \mu = 0,
  \eqn \condC
$$
with the constraint $\int \phi_0 = 1$.
The bosonic potential $w$ is in general chosen to depend on a
coupling constant $g$ in such a way that $w(x/\sqrt g) =
\tilde w(x)/g$, where the rescaled potential $\tilde w$ has no
explicit
dependence on $g$.  Rescaling $x\to x/\sqrt g$, $\phi\to \phi/\sqrt
g$ and $\mu \to \mu/g$, all explicit dependence on $g$ scales
out of the stationarity condition, which becomes
$$
  {\pi^2\over 2}\, \phi_0^2(x)
      + \tilde w(x)  - \mu = 0,
  \eqn \condD
$$
while the constraint becomes $\int \phi_0 = g$.  We see that by
changing the coupling constant $g$, which now appears in the
normalization condition, we can freely adjust
the fermi level $\mu$ with respect to the rescaled potential
$\tilde w$.  In particular, as $\mu$ approaches a $-x^2$ maximum, the
time of flight $L \propto \int dx/\phi_0$ becomes infinite.

Let us now try to duplicate this in the fermionic case.
Taking
the potential in \condB\ to have the general scaling $v(g^a x) = g^b
\tilde v(x)$, where $\tilde v$ is independent of $g$, and rescaling
$x\to g^ax$, $\phi\to g^c\phi$ and $\mu\to g^{2c}\mu$, the
stationarity condition \condB\ becomes
$$
  g^{2c}\,{\pi^2\over 2}\, \phi_0^2(x)
      + g^{2b}\, \half\, \tilde v^2(x)  - g^{2c}\mu
      - g^{b+c}\pv dy \,{\tilde v(x)-\tilde v(y)\over
x-y}\,\phi_0(y) = 0,
  \eqno\eq
$$
while the constraint becomes $\int \phi_0 = g^{-a-c}$.
Expanding $\tilde v$ about $x$ and using this constraint, the
stationarity condition becomes
$$
\eqalign{
  &
  0 = g^{2c}\,{\pi^2\over 2}\, \phi_0^2(x)
      + g^{2b}\, \half\, \tilde v^2(x)  - g^{2c}\mu  \cr
  &\quad
      - g^{b+c-a-c} \,\tilde v' (x)
      + \half\,g^{b+c-a-c}\, \tilde v'' (x)\,x
      - \half\,g^{b+c}\, \tilde v'' (x) \int dy\,y\, \phi_0(y)  \cr
  &\quad   + \dots.  \cr
}  \eqno\eq
$$
We see that if we choose $a = -b =-c$, an overall factor $g^{b+c}$
divides out of the equation and we find
$$
\eqalign{
  &
  0 = {\pi^2\over 2}\, \phi_0^2(x)
      +  \half\, \tilde v^2(x)  - \mu  \cr
  &\quad
      - \tilde v' (x)
      + \half\ \tilde v'' (x)\,x
      - \half\, \tilde v'' (x) \int dy\,y\, \phi_0(y)  \cr
  &\quad   + \dots,  \cr
}  \eqno\eq
$$
while the constraint is given by $\int\phi_0 = g^{-a-c} = 1$.  We have
thus succeeded in getting rid of all explicit and implicit dependence
of $\phi_0$ on $g$.  Rewriting this we therefore get
$$
  {\pi^2\over 2}\, \phi_0^2(x)
      + \half\, \tilde v^2(x)  - \mu
      - \pv dy \,{\tilde v(x)- \tilde v(y)\over x-y}\,\phi_0(y) = 0,
  \eqno \eq
$$
with the constraint $\int\phi_0 = 1$.  A point we want to
emphasize is that in contrast to the bosonic case the parameter $g$
does not appear
in the constraint, so that the effective fermi level is not a free
parameter relative to the rescaled effective potential in \? and
cannot be adjusted to approach a maximum.  (The effective fermi
level here is given by $\mu$ plus any $x$-independent terms
arising from the cross term in \?).  The time of flight,
given by $L \propto \int dx/\phi_0$, remains finite and no true
extra dimension is generated.

The above argument assumes that $v(x)$ depends in a particularly
simple way on only one parameter $g$,
and does not make a statement on the
possibility or not of reproducing an analogue of the bosonic scaling
limit when this is not the case.  This seems very hard to do in
specific examples that have been considered (not reproduced here),
but more work needs to be done to gain a better understanding of this
issue.

In conclusion, the scaling behaviour of the supersymmetric theory,
when expanded about a classical vacuum, corresponds to that of
$d=0$ matrix models.  This is in contrast to the bosonic collective
field
theory and no true extra dimension is generated in the
scaling limit.

In view of this fact, and also considering the
argument of section (4.5), where it was shown that the
turning
point regularizations of three point functions that have been used in
the
bosonic collective field theory are incompatible with supersymmetry,
the true structure of any $d=1$ supersymmetric string theory
 remains very much a mystery.

One clue towards finding the correct structure of such a theory
may be the fact that in the
bosonic $d=1$ string a W-algebra structure arises naturally [4,
5] --- the idea would therefore be to investigate theories that
naturally fit into suitable supersymmetric generalizations of the
W-algebra [35].  This issue is currently under investigation.

\endpage

\chapter {CONCLUSIONS}

\noindent
In this thesis an investigation was made of the super-Calogero model
with particular emphasis on its continuum formulation and possible
application in the context of supersymmetrizing the bosonic collective
$d=1$ string field theory.

We started with a discussion of the model both in a discrete and
continuum formulation, demonstrating its equivalence with the
Jevicki-Rodrigues supersymmetrized collective field theory and the
Marinari-Parisi supersymmetric matrix model.  Upon quantization, the
continuum fields were found to have nontrivial commutation relations
involving square roots of the density field, which led to an infinite
sequence of higher order vertices in the perturbative Hamiltonian as
well as the supersymmetry generators.  This was in contrast to the
bosonic collective field theory, where only a cubic
vertex is required.

We then discussed the potential free case, with an explicit
calculation of the spectrum and the cubic vertices, in the process
identifying the superpartner of the masless tachyon to be a massless
Majorana fermion.  We compared a particular spacetime
formulation of the theory with gauge fixed two-dimensional
supergravity and
found that the comparison breaks down because of nonlocal terms in the
interactions.  The exact spectrum, as previously obtained in the
bosonic collective field theory, was generalised to the supersymmetric
case and a concise representation of the superpartners of the bosonic
eigenstates was obtained.  In addition to this, a formulation was
postulated in which to investigate nonperturbative effects such as
solitons in a semiclassical analysis.

For the harmonic case, the semiclassical spectrum was obtained and
found to be similar to that of the free case.  The cubic vertices were
then calculated, an undertaking that turned out to be very sensitive
to the method of regularization.  If one wants to preserve
supersymmetry, it seems that one cannot generate a bosonic sector
compatible with the bosonic collective field theory (no $a_m a_n
a_p$ terms).

We then moved on to more general potentials, first showing that the
semiclassical spectrum is always supersymmetric provided one expands
around a supersymmetric classical configuration.  For a specific class
of superpotentials we then demonstrated that the semiclassical
spectrum is that of a massless scalar boson and a massless Majorana
fermion, the quadratic terms taking a particularly simple form in
terms of the scaling coordinates.

We concluded with a discussion of the scaling properties of the
superpotentials and the possibility of a spacetime interpretation of
the double scaled theory.  It was argued that
the scaling poperties of the classical vacuum density
correspond to that found in zero-dimensional $c<1$ models, where no
true extra dimension is created.
This is in contrast to the bosonic case; thus the spacetime
interpretation remains problematic and more work is needed on this
issue.

At this point various directions of study may be pursued.
On the one hand, in view of the fact that in the
bosonic case a W-algebra structure has been identified [4, 5], one can
attempt to associate to the above theory a suitable supersymmetric
generalization of the W-algebra [35].  One could also start
from the opposite direction: the true
theory of $d=1$ superstrings may not correspond to the model
described in this thesis, in which case one may find
clues to the correct description directly from the structure of
suitable supersymmetric extensions of the $W$-algebra.
Finally,
there is a lot to be done to extend existing results for the discrete
Calogero model [34] to its supersymmetric generalization.

\endpage

\centerline{\bf REFERENCES}
\bigskip
\pointbegin
    A. Jevicki and B. Sakita, {\it Nucl. Phys.} {\bf B165} (1980) 511.
\point
    S.R. Das and A. Jevicki, {\it Mod. Phys. Lett.} {\bf A5} (1990)
    1639.
\point
    J. Polchinski, {\it Nucl. Phys.} {\bf B346} (1990) 253.
\point
    J. Avan and A. Jevicki, Brown University preprints,
        BROWN-HET-801 (1991), BROWN-HET-824 (1991) and BROWN-HET-839
       (1991);
\point
    J. Polchinski, University of Texas preprint UTTG-06-91 (1991);
    D. Minic, J. Polchinski and Z. Yang, University of Texas preprint
    UTTG-16-91;
    E. Witten, \lq\lq Ground ring of two dimensional string
     theory\rq\rq, IASSNS-HEP-91/51
     (1991), to appear in {\it Nucl. Phys. B.};
    E. Witten and B. Zwiebach, \lq\lq Algebraic structures and
    differential geometry in 2d string theory\rq\rq,
    SLAC-PUB-IASSNS-HEP-92/4, MIT-CTP-2057 (1992).
\point
    K. Demeterfi, A. Jevicki and J.P. Rodrigues,
        {\it Nucl. Phys.} {\bf B362} (1991) 173;
        K. Demeterfi, A. Jevicki and J.P. Rodrigues,
        {\it Nucl. Phys.} {\bf B365} (1991) 499.
\point
    K. Demeterfi, A. Jevicki and J.P. Rodrigues,
    {\it Mod. Phys. Lett.} {\bf A6} (1991) 3199.
\point
    D. J. Gross and N. Miljkovic, {\it Phys. Lett.} {\bf B238} (1990)
        217;
        E. Br\'ezin, V. A. Kazakov and Al. B. Zamolodchikov,
        {\it Nucl. Phys.} {\bf B338} (1990) 673;
        P. Ginsparg and J. Zinn-Justin, {\it Phys. Lett.} {\bf B240}
        (1990) 333.
 \point
    P. Di Francesco and D. Kutasov, {\it Phys. Lett.} {\bf B261}
      (1991)
        385;
        N. Sakai and Y. Tanii, Tokyo Inst. of Technology preprint
        TIT/HEP-1684;
        Y. Kitazawa, Harvard preprint HUTP-91/A034 (1991);
        M. Bershadsky and I. R. Klebanov, {\it Phys. Rev. Lett.} {\bf
        65} (1990) 3088;
        Princeton preprint PUPT-1197 (1990);
\point
    D. J. Gross and I. R. Klebanov,
        {\it Nucl. Phys.} {\bf B352} (1991) 671, {\it B356} (1991) 3;
        A. Sengupta and S. Wadia, {\it Int. J. Mod. Phys.} {\bf A6}
        (1991) 1961.
\point
    G. Moore, Rutgers and Yale preprint RU-91-12 and YCTP-P8-91
    (1991).
\point
    D. Kutasov and N. Seiberg, {\it Phys. Lett.} {\bf B251} (1990) 67.
\point
    J. Alfaro and P. H. Damgaard, {\it Phys. Lett.} {\bf B222} (1989)
        425;
        E. Marinari and G. Parisi, {\it Phys. Lett.} {\bf B240} (1990)
        375;
        M. Karliner and A. A. Migdal, {\it Mod. Phys. Lett.} {\bf A6}
        (1990) 2565;
        S. Belluci, T.R. Govindrajan, A. Kumar and R. N. Oerter,
        {\it Phys. Lett.} {\bf B249} (1990) 49;
        J. Gonzalez, {\it Phys. Lett.} {\bf B255} (1991) 367;
        L. Alvarez-Gaume' and J. L. Manes, CERN preprint CERN-TH-6067/91;
        A. Dabholkar, Rutgers University preprint RU-91-20 (1991).

\point
    E. Abdalla, M.C.B. Abdalla and D. Dalmazi, K. Haradi,
        \lq\lq Correlation functions in super Liouville theory\rq\rq,
        S\~ao Paolo preprint (1991);
        K. Aoki and E. d'Hoker, \lq\lq Correlation functions of
        minimal models
        coupled to two dimensional quantum supergravity\rq\rq,
        UCLA preprint UCLA/91/TEP/33 (1991);
        L. Alvarez-Gaume' and P. Zaugg, \lq\lq Some correlation
        functions of
        minimal superconformal models coupled to supergravity\rq\rq,
        CERN preprint CERN-TH.6243/91 (1991).
\point
    P. Di Francesco and D. Kutasov, Princeton University preprint
        PUPT-1276 (1991);
    P. Bouwknegt, J. McCarthy and K. Pilch, \lq\lq Ground ring
    for the 2d NSR string\rq\rq, CERN preprint CERN-TH.6346/91 (1991),
   to appear in {\it Commun. Math. Phys};
   K. Itoh and N. Ohta, \lq\lq Spectrum of two-dimensional
   (super)gravity\rq\rq,  Osaka University and Brown University
   preprints  OS-GE 22-91, BROWN-HET-844.
\point
    A. Jevicki and J.P. Rodrigues, {\it Phys. Lett.} {\bf B268} (1991)
    53.
\point
    A. Jevicki, \lq\lq Nonperturbative collective field theory\rq\rq,
    Brown University preprint BROWN-HET-807 (1991).
\point
    E. Br\'ezin, C. Itzykson, G. Parisi and J.B. Zuber, {\it Commun.
    Math. Phys.} {\bf 59}, 35-51 (1978).
\point
    J.P. Rodrigues, {\it Phys. Rev.} {\bf D26} (1982) 2833.
\point
    M. Casartelli, G. Marchesini and E. Onofri, {\it J. Phys.} {\bf
    A13} (1980) 1217.
\point
    A. Jevicki and H. Levine, {\it Phys. Rev. Lett.} {\bf 44} (1980)
    1443.
\point
    R. Calogero, {\it J. Math. Phys.} {\bf 12} (1971) 419.
\point
    D.Z. Freedman and P.F. Mende, {\it Nucl. Phys.} {\bf B344} (1990)
    317.
\point
    I.R. Klebanov and A.M. Polyakov, {\it Mod. Phys. Lett.} {\bf 35A}
    (1991) 3273.
\point
    M.B. Green, J.H. Schwarz and E. Witten, \lq\lq Superstring
    Theory\rq\rq, Volume I, Cambridge University Press (1987).
\point
   P.A.M. Dirac, \lq\lq Lectures on Quantum Mechanics\rq\rq,
   Yeshiva University press (1964).
\point
   F.A. Berezin, \lq\lq The Method of Second Quantization\rq\rq,
   Academic Press (1966).
\point
    A. D'Adda, \lq\lq Comments on supersymmetric vector and matrix
    models\rq\rq, Turin University preprint DFTT 24/91 (1991);
    L. Alvarez-Gaum\'e and J.L. Ma\~nes, \lq\lq Supermatrix
    models\rq\rq, CERN preprint CERN-TH.6067/91 (1991).
\point
    E. Witten, {\it Nucl. Phys.} {\bf B185} (1981) 513,
    {\it Nucl. Phys.} {\bf B202} (1982) 253;
    P. Salomonson and J.W. van Holten, {\it Nucl. Phys.} {\bf B196}
       (1982) 509;
    D. Lancaster, {\it Nuovo Cim.} {\bf 79A} (1984) 28.
\point
    I. Bars, {\it J. Math. Phys.} {\bf 21} (1980) 2678;
    T. Maekawa, {\it J. Math. Phys.} {\bf 26} (1985) 1902.
\point
   D.J. Gross and A.A. Migdal, {\it Phys. Rev. Lett.} {\bf 64} (1990)
   127;  M.R. Douglas and S. Shenker, {\it Nucl. Phys.} {\bf B335}
   (1990)
   635;  E. Br\'ezin and V. Kazakov, {\it Phys. Lett.} {\bf B236}
   (1990)
   144;  C. Crnkovi\'c, P. Ginsparg and G. Moore, {\it Phys. Lett.}
   {\bf
   B237} (1990); D.J. Gross and A. Migdal, {\it Phys. Rev. Lett.} {\bf
   64} (1990) 717;  E. Br\'ezin, M.R. Douglas, V.A. Kazakov and S.H.
   Shenker, {\it Phys. Lett.} {\bf B237} (1990) 43; M. Douglas, {\it
   Phys. Lett.} {\bf 238B} (1990) 176;  K. Demeterfi, N. Deo, S. Jain
   and Chung-I Tan, {\it Phys. Rev.} {\bf D42} (1990) 4105.
\point
   J. Feinberg, \lq\lq String field theory for $d\le 0$ matrix
   models via Marinari-Parisi\rq\rq, Israel Inst. of Technology
   preprint TECHNION-PH-92-1 (1992);
   J.D. Cohn and H. Dykstra, \lq\lq The Marinari-Parisi model
   and collective field theory\rq\rq,  Fermilab preprint
   FERMILAB-PUB-92/35-T (1992).
\point
   B. Sazdovi\'c, private communication.
\point
  F. Calogero, {\it J. Math. Phys.} {\bf 10} (1969) 2191,
  {\it J. Math. Phys.} {\bf 10} (1969) 2197,
  {\it J. Math. Phys.} {\bf 12} (1971) 419,
  {\it Lett. Nuovo Cim.} {\bf 19} (1977) 505,
  {\it Lett. Nuovo Cim.} {\bf 20} (1977) 251;
  A.M. Perelomov, {\it Teor. Mat. Fiz.} {\bf 12} (1971) 364,
  {\it Ann. Inst. Henri Poincar\'e} {\bf 28} (1978) 407;
  M.A. Olshanetsky and A.M. Perelomov, {\it Phys. Rep.} {\bf 71}
  (1981) 313, {\it Phys. Rep.} {\bf 94} (1983) 313;
  M. Bruschi and F. Calogero, {\it Lett. Nuovo Cim.} {\bf 24} (1879)
  601;
  S. Ahmed, M. Bruschi, F. Calogero, M. Olshanetsky and A.M.
  Perelomov, {\it Nuovo Cim.} {\bf B49} (1979) 173.
\point
  E. Bergshoeff, B. de Wit and M. Vasiliev, \lq\lq The structure of
  the
  super-$W_\infty$ algebra\rq\rq, CERN preprint CERN-TH.6021/91;
  E. Bergshoeff, C.N. Pope, L.J. Romans, E. Sezgin and X. Shen,
  {\it Phys. Lett.} {\bf B245} (1990) 447;
  T. Inami, Y. Matsuo and I. Yamanaka, {\it Phys. Lett.}
  {\bf B215} (1988) 701;
  F. Yu, \lq\lq The super-KP origin of the super-$W_{1+\infty}$
  algebra and its topological version\rq\rq, Utah University preprint
  UU-HEP-91/12 (1991).
\point
  B. Zwiebach, {\it Ann. Phys.} {\bf 186} (1988) 111.
\point
  J. Cohn and S.P. DeAlwis, IAS preprint IASSNS-HEP-91/7 (1991).

\vfill
\endpage

\end